%% file: main.tex
\documentclass[sigconf]{acmart}
\usepackage{soul}
\usepackage{graphicx}        
\usepackage{listings}
\usepackage{multirow}
\usepackage{enumitem}

\copyrightyear{2025} \acmYear{2025} \setcopyright{acmlicensed}\acmConference[CHI '25]{CHI Conference on Human Factors in Computing Systems}{April 26-May 1, 2025}{Yokohama, Japan} \acmBooktitle{CHI Conference on Human Factors in Computing Systems (CHI '25), April 26-May 1, 2025, Yokohama, Japan} \acmDOI{10.1145/3706598.3713785} \acmISBN{979-8-4007-1394-1/25/04}

\begin{document}

\title{Dancing With Chains: Ideating Under Constraints With UIDEC in UI/UX Design}

\author{Atefeh Shokrizadeh}
\email{atefeh.shokrizadeh@polymtl.ca}
\orcid{0009-0004-0743-6875}
\affiliation{%
  \institution{Polytechnique Montreal}
  \city{Montreal}
  \state{Quebec}
  \country{Canada}
}

\author{Boniface Bahati Tadjuidje}
\email{bahati-tadjuidje.boniface@polymtl.ca}
\orcid{0009-0004-8386-9664}
\affiliation{%
  \institution{Polytechnique Montreal}
  \city{Montreal}
  \state{Quebec}
  \country{Canada}
}

\author{Shivam Kumar}
\email{shivam.sng14@gmail.com}
\orcid{0009-0008-2624-2813}
\affiliation{%
  \institution{Polytechnique Montreal}
  \city{Montreal}
  \state{Quebec}
  \country{Canada}
}

\author{Sohan Kamble}
\email{sohan.kamble21@gmail.com}
\orcid{0009-0002-7109-1969}
\affiliation{%
  \institution{Polytechnique Montreal}
  \city{Montreal}
  \state{Quebec}
  \country{Canada}
}

\author{Jinghui Cheng}
\email{jinghui.cheng@polymtl.ca}
\orcid{0000-0002-8474-5290}
\affiliation{%
  \institution{Polytechnique Montreal}
  \city{Montreal}
  \state{Quebec}
  \country{Canada}
}

\begin{abstract}
UI/UX designers often work under constraints like brand identity, design norms, and industry guidelines. How these constraints impact designers' ideation and exploration processes should be addressed in creativity-support tools for design. Through an exploratory interview study, we identified three designer personas with varying views on having constraints in the ideation process, which guided the creation of UIDEC, a GenAI-powered tool for supporting creativity under constraints. UIDEC allows designers to specify project details, such as purpose, target audience, industry, and design styles, based on which it generates diverse design examples that adhere to these constraints, with minimal need to write prompts. In a user evaluation involving designers representing the identified personas, participants found UIDEC compatible with their existing ideation process and useful for creative inspiration, especially when starting new projects. Our work provides design implications to AI-powered tools that integrate constraints during UI/UX design ideation to support creativity.
\end{abstract}

\ccsdesc[500]{Human-centered computing~User interface design}

\keywords{User Interface Design, Constraint, Inspiration, Ideation, Creativity Support}

\begin{teaserfigure}
    \includegraphics[width=\textwidth]{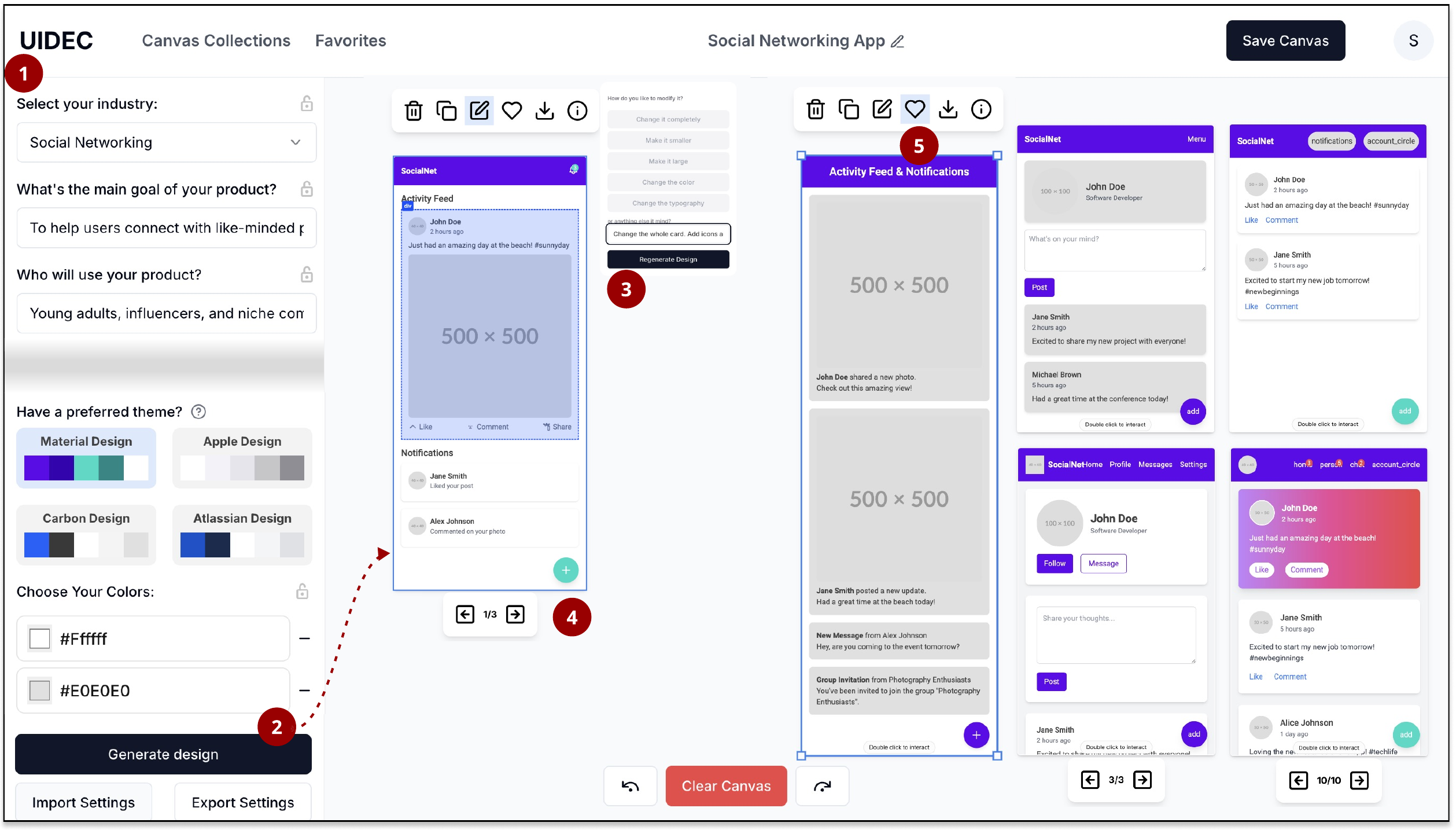}
    \caption{UIDEC allows UI/UX designers to specify design constraints \textcircled{1}, generate and organize design ideas on a canvas \textcircled{2}, refine or regenerate specific UI elements \textcircled{3}, navigate through different generated versions  \textcircled{4}, and save and arrange design examples \textcircled{5}, among other features. We investigated the design of UIDEC to understand designers' needs in creative ideation under constraints and to identify opportunities for using AI-powered tools to support this process.}
    \label{fig:teaser}
    \Description{This figure highlights the main features of UIDEC, including (1) specifying design constraints, (2) generating and organizing design ideas on a canvas, (3) refining or regenerating specific UI elements, (4) navigating through different generated versions, and (5) saving and arranging design examples.}
\end{teaserfigure}

\maketitle

\input{s_introduction}
\input{s_relatedwork}
\input{s_formativestudy}
\input{s_systemdesign}
\input{s_methods}
\input{s_results}
\input{s_discussion}

\begin{acks}
We thank our participants for their time and valuable insights. We also thank the anonymous reviewers for helping us improve the paper. This work is partially supported by the Canada Research Chairs program (CRC-2021-00076), Fonds de recherche du Québec -- Nature et technologies (2022-PR-299099), and the Natural Sciences and Engineering Research Council of Canada (RGPIN-2018-04470).
\end{acks}
\balance

\bibliographystyle{ACM-Reference-Format}
\bibliography{references}

\appendix
\input{s_appendices}

\end{document}

%% file: s_introduction.tex
\section{Introduction}
UI/UX designers take on daily challenges in creating effective and usable design work. On one hand, they strive to cultivate innovative designs that set their product apart in the fiercely competitive landscape of software applications and services. On the other hand, their designs are often constrained by specific requirements, clients' preferences, and business considerations of the target product~\cite{wood1997user}. Thus, UI/UX designers frequently find themselves engaged in a delicate dance in which they seek to unleash their creativity under imposed constraints.

Drawing inspiration from existing \textit{image-based design examples} (e.g., screenshots, UI mockups, and layouts) plays an essential role in this creative process~\cite{Goncalves2014}. This practice not only ensures that important design conventions are followed, but it also reinforces crucial creativity mechanisms when transforming, combining, and adapting elements from previous design ideas~\cite{Eckert2000, Swearngin2018}. Successful design innovations often originate from such an analogy-based inspiration mechanism enabled by design examples. There are currently many online platforms that facilitate the sharing and searching of image-based UI design examples for inspiration (e.g., dribbble.com, behance.net, and siteinspire.com). Several research studies also proposed methods aimed at recommending or generating design examples (e.g.,~\cite{mozaffari2022ganspiration,huang2019swire}). Previous studies established that constraints can inspire creative thinking~\cite{Acar2019, stokes2006creativity}. However, the current tools and approaches fail to incorporate the UI/UX design requirements and restrictions during the process of retrieval, recommendation, and generation of design examples; therefore, they cannot effectively address the common ``creativity-under-constraint'' challenge faced by designers.

In this work, we targeted this challenge and specifically investigated how creativity-support tools can facilitate design ideation under constraints. To this end, we first conducted an exploratory interview study with experienced designers to understand how they worked with constraints in practice. The results revealed that the most common constraints included user characteristics, industry standards, design systems, technical feasibility, brand identity, and business needs. Moreover, designers perceived the impact of these constraints on creativity differently, considering them as either limitations or ideation materials and guidelines. Based on the study findings, we created three user personas and five design considerations, which informed the design of UIDEC, which stands for \textbf{UI} \textbf{D}esign \textbf{E}xploration under \textbf{C}onstraints. UIDEC, leveraging generative AI technologies, allows UI/UX designers to specify their constraints and generate design examples accordingly, with minimal need for prompt writing. Designers can further use UIDEC to iterate on generated designs and collect mood boards of their favorite ideas (see Figure~\ref{fig:teaser}). Subsequently, we conducted a user study with ten UI/UX designers representing our three personas to evaluate the tool. The study participants found that ideating with UIDEC minimized irrelevant design exploration and enhanced the efficiency and effectiveness of the ideation process. Participants' feedback allowed us to reflect on the design considerations we created and provide further implications for creating future tools to better streamline the UI/UX design ideation process. Overall, our investigation with UIDEC contributed to the understanding of designers' multifaceted needs when ideating under constraints and highlighted important design considerations for informing future AI-based tools to support this creative activity.

%% file: s_relatedwork.tex
\section{Related Work}
Our work builds upon prior research on the impact of constraints on design creativity, tools that support design inspiration, and generative AI for UI/UX design. We briefly discuss each of these areas below.

\subsection{Constraints and Creativity in Design}
The effects of constraints on creativity have been carefully investigated in the literature. \citet{stokes2006creativity} established that constraints guide practitioners, including designers, toward novel solutions by defining focused search spaces where creators can systematically explore alternatives. Constraints help define the problem space, supporting beginners in structuring the solutions and encouraging experts to frame new design goals to achieve innovation~\cite{stokes2006creativity}. Building on this work, \citet{stokes2014crossing} later developed a model that demonstrated how ``paired constraints'' -- one specifying solutions and the other restricting them -- direct ideation through a process of substitution and iteration, leading to more innovative outcomes. This understanding of constraints helps explain how designers can systematically navigate the solution space while maintaining creative freedom. Building upon this notion, \citet{Biskjaer2014Decisive} argued that during creative practices, voluntary self-binding through so-called ``decisive constraints'' that radically limit the solution space may in fact accelerate the innovation process.

The process of constraint-based creative engagement was explored in various fields within and beyond design. For example, \citet{dahl2007thinking} focused on regular consumers of products that offer creative opportunities and found that creative tasks balancing autonomy with competence-building guidance lead to greater engagement and enjoyment. Focusing on creativity in organizations, \citet{Acar2019} also synthesized a taxonomy of constraints and suggested an ``inverted U-shaped relationship'' between constraints and creativity; that is to say, ``the right amount'' of constraints unlocks and stimulates creativity. \citet{Onarheim2012} reported on a case study in the engineering design of medical equipment and found that designers were highly constraint-focused and adopted various strategies to manipulate constraints to achieve their creative goals and generate novel solutions. Related to design in HCI, \citet{Biskjaer2014} argued that the notion of \textit{design space} should be understood as a constraint-based conceptual space, which designers co-constitute, explore, and shape iteratively throughout the creative design process. These previous efforts highlighted the importance of incorporating the consideration of constraints when navigating and facilitating creative processes, including UI/UX design. However, how to support this process for UI/UX designers remains unexplored in the literature. Our work addresses this gap by investigating UI/UX designers' specific needs when ideating under constraints, as well as considerations for tool design to support these activities.

\subsection{Tools for Design Inspiration}
Design inspiration plays a crucial role in the UI/UX design process, often involving the exploration of existing designs and the generation of novel ideas~\cite{sharmin2009understanding}. In practice, professional designers rely on platforms like Dribbble\footnote{https://dribbble.com}, Behance\footnote{https://www.behance.net}, and Siteinspire\footnote{https://www.siteinspire.com} for discovering design examples. In the literature, many previous works have explored Creativity Support Tools (CSTs) to facilitate design inspiration, with a particular focus on supporting ideation and exploration~\cite{frich2019mapping}. We briefly review this literature below.

One type of those tools is focused on leveraging automated methods for retrieving and managing image-based design examples for inspiration. For example, \citet{Lee2010} proposed a tool that suggests to web designers image-based examples similar to their current design work. Following a similar approach, \citet{ritchie2011dtour} proposed a design exploration tool that allows querying design examples by descriptive text, including color keywords or style terms. \citet{huang2019swire} also introduced a technique based on a deep convolutional neural network (CNN), trained with an open dataset, for retrieving UI design images based on hand-drawn sketches. 

Adaptation tools have further expanded these capabilities through various innovative approaches. For example, \citet{swearngin2018rewire} introduced Rewire for retargeting web designs across different layouts while preserving their original visual style and structure. \citet{zhong2021spacewalker} developed SpaceWalker for exploring design alternatives through systematic layout space navigation, enabling designers to discover new variations while maintaining design coherence. \citet{swearngin2020scout} advanced these concepts further by providing tools for manipulating existing designs through direct design synthesis and adaptation techniques. More recently, \citet{son2024genquery} proposed GenQuery, a tool that supports visual-based iterative design search and generation for inspiration. These tools demonstrated the evolution from simple retrieval to sophisticated manipulation and adaptation of design examples.

Mood boards are another common type of CST tool for visual inspiration in design processes. For example, \citet{koch2020semanticcollage} developed SemanticCollage, which uses AI-recognized semantic labels to support both visual and semantic-based ideation in digital mood boards. Building on this, ImageSense~\cite{koch2020imagesense} integrated AI-suggested images in a collaborative mood board setting. While these tools demonstrate the potential of AI in supporting design inspiration, they primarily focus on organizing and suggesting existing images rather than generating new designs.

UIDEC extends this line of work by integrating generative AI capabilities directly into the design inspiration process, allowing designers to not only explore existing designs but also generate and iterate on novel UI concepts.

\subsection{Generative Models for UI Creation}
Recent advancements in generative AI have opened new possibilities for UI generation. While early approaches like LayoutGAN~\cite{li2019layoutgan} focused on generating layouts from random noise, more recent methods have explored ways to incorporate project and design considerations during the generation process. For example, the Layout Generative Network~\cite{zheng2019content} introduced content-aware layout generation, LayoutFlow \cite{guerreiro2024layoutflow} used flow matching, Spot the Error \cite{lin2024spot} explored non-autoregressive methods, and the Retrieval-Augmented Layout Transformer \cite{horita2023retrieval} leveraged nearest neighbor layout examples to improve generation quality. Beyond layout generation, \citet{zhao2021guigan} proposed the GUIGAN method to compose full-fledged user interfaces with existing components extracted from a dataset. While being able to generate realistic GUIs, this approach is not able to produce new styles that have potential inspirational power, since the generated GUIs are just a combination of other GUIs from their dataset. To address this issue, \citet{mozaffari2022ganspiration} proposed GANSpiration, which uses StyleGAN to merge the styles of the source UI with random images to generate new UI images for inspiration.

More recently, approaches leveraging large language models (LLMs) have shown promise in UI generation. For example, LayoutNUWA~\cite{tang2023layoutnuwa} reformulates layout elements into HTML code, taking advantage of the rich design knowledge embedded in LLMs. \citet{cheng2024graphic} also introduced Graphist, which uses a large multimodal model for Hierarchical Layout Generation (HLG). Building on these ideas, our tool UIDEC uses a multimodal LLM to generate HTML artifacts based on natural language constraints and reference images. Commercially, AI-based tools such as Uizard\footnote{https://uizard.io} and Galileo\footnote{https://www.usegalileo.ai}, as well as AI-enhanced features in design tools like Figma and Framer, are also becoming available to generate prototypes. However, most of these LLM-based tools rely on a natural language input that requires extensive prompt engineering, which presumably is a suboptimal interface for design creativity support~\cite{Morris2024}.

As AI tools become increasingly prevalent in design workflows, effective human-AI collaboration has emerged as a crucial area of study. For example, \citet{tholander2023design} investigated how designers integrated generative AI models into their ideation processes, emphasizing the importance of maintaining designer control while leveraging AI capabilities. Through DesignPrompt, \citet{peng2024designprompt} further demonstrated how multimodal interaction with generative AI can support more intuitive and expressive design exploration. Interestingly, \citet{son2024genquery} found that the unpredictability of generative AI can stimulate creativity in design processes. However, \citet{anderson2024homogenization} warns that AI tools can lead to output homogenization across users, emphasizing the need for mechanisms to increase variations for creative exploration. These previous efforts suggested that effective human-AI collaboration in design is not just about control, but also about leveraging AI's potential for unexpected inspiration. UIDEC builds on these insights by addressing the identified challenges and balancing designer control with AI-driven inspiration.

%% file: s_formativestudy.tex
\section{Exploratory Interview Study}
We conducted an exploratory interview study with experienced UI/UX designers to gain insights into how design constraints impact the exploration and ideation process and how AI can assist designers in generating ideas. Our goal was to explore three key research questions: (1) What types of constraints do designers face throughout the design process? (2) How do these constraints influence their ability to generate new ideas? (3) What features would they like to see in an AI-powered tool that assists design exploration?

\subsection{Interview Methods}
Participants were recruited through personal connections, advertisements on LinkedIn, and targeted emails to professional designers. A total of 19 UI/UX designers participated in the study, representing a diverse range of experience levels, from recent graduates to professionals with up to 15 years of experience in the field. The participants included designers working in large teams within companies, design team leaders, and freelancers. Table~\ref{tab:participant-info-exploratory} summarizes the characteristics of our participants.

\begin{table}[t]
\centering
\small
\caption{Summary of characteristics of participants of the interview study}
\label{tab:participant-info-exploratory}
\begin{tabular}{cclc}
\toprule
\textbf{ID} & \textbf{Gender} & \textbf{Role/Profession} & \textbf{Years of Exp.} \\ \midrule
P1                      & Female          & UI/UX Designer            & 5 years                \\
P2                      & Male            & Product Design Lead       & 10 years                \\
P3                      & Male            & Product Design Lead            & 8 years                  \\
P4                      & Female          & UI/UX Designer            & 4 years              \\
P5                      & Female          & UX Designer               & 6 years                 \\
P6                      & Female          & UX Designer               & 2 years                 \\
P7                      & Male            & Product Designer          & 2 years                 \\
P8                      & Female          & Graphic Designer/UX/UI Designer & 10 years         \\
P9                      & Male          & Senior product designer               & 8 years              \\
P10                     & Female          & Interaction Design Student & 2 years                  \\
P11                     & Male            & UX Designer/University Lecturer       & 15 years                 \\
P12                     & Female          & UX Junior Consultant      & 2 years                  \\
P13                     & Male            & Product Designer          & 4 years                  \\
P14                     & Female          & UI/UX Designer            & 4 years              \\
P15                     & Female            & UI/UX Designer            & 2 years                  \\
P16                     & Male            & UI/UX Designer                   & 2 years              \\
P17                     & Male            & Graphic Designer            & 2 years                 \\ 
P18                     & Male            & UI/UX Designer              & 6 years              \\
P19                     & Male            & UI/UX Designer              & 2 years               \\
\bottomrule
\end{tabular}
\end{table}

We conducted semi-structured interviews via Zoom with each participant. Each session, lasting approximately 45 minutes, was recorded with the participant's consent; each participant was compensated with \$30 CAD. The interview protocol began with questions regarding the participants' backgrounds in UI/UX design, the projects they have been involved in, and their typical design workflows. Following this, we delved into the constraints they encountered during ideation, examining how these constraints affected their creative processes. Finally, we discussed their current use of AI tools in design and explored their preferences for features in an ideal AI tool. 

Data analysis began with a verbatim transcription of all recorded interviews. We then conducted thematic analysis~\cite{Vaismoradi2013, aronson_pragmatic_1995} collaboratively on the transcribed interview data. Inductive coding was initially performed. Codes were then grouped iteratively into broader themes to answer each research question.

\subsection{Interview Results}

\subsubsection{Constraints encountered by designers.}
\label{sec:interview_constraints}
When participants were asked about ``constraints,'' their initial thoughts centered around resources like time and budget. These \textbf{Resource-Based Constraints} often shaped the scope and approach of their design processes. For example, P3 explained: ``\textit{I go into solutioning with the impact of the scope, ... the impact of what the budget is and time and stuff like that. With these, I would have a better understanding of what is possible and what is not.}'' Similarly, P11 highlighted, ``\textit{Sometimes it's not possible to do everything you want because for example there's a cost to develop a custom component}''. As the conversation progressed, participants began to discuss \textbf{Content-Based Constraints}, which is the focus of our study and can be categorized as follows:

\textit{(a) User Characteristics:} Naturally, designers must consider user characteristics to, for example, use familiar patterns and structures to avoid adding cognitive load. For example, P5 discussed how this type of constraint can challenge innovation: ``\textit{It's very difficult to get [a new] idea out because people are not used to it. People have not seen it, so they do not know how it is actually going to be working out there. Because, you know, as people, we are very habitual... And that is a huge consideration.}''

\textit{(b) Industry Norms:} Designers must also adhere to norms specific to the industry that they design for. For instance, designing for the food industry differs significantly from designing for health applications. For example, P2 noted ``\textit{There are certain colors that lend themselves better to certain things... red and orange... are connected usually to food and appetite. So there are general, you know, categories.}'' P12 also discussed balancing industry norms and uniqueness of the product: ``\textit{So it was mostly like trying to be part of the same market of people, like other companies. But I did use a bit of different colors to not look exactly like Netflix or Prime.}''

\textit{(c) Design Systems:} Another key factor is the design system in use. Some companies have internal design systems that designers must follow, with limited flexibility for modifications. As P11 mentioned ``\textit{[The design system] is not always up to me to decide. Sometimes it already has been decided, and sometimes it needs to be created.}'' Every so often, designers must consult with the development team regarding the design system and adhere to it. As P9 stated, ``\textit{That usually depends on the code base the development team is working with. So there's less friction when it comes to the design handoff.}''

\textit{(d) Technical Feasibility:} Designers also face technical constraints that require collaboration with the development team to assess the feasibility of their ideas. For instance, P5 described ``\textit{As designers, we tend to go overboard with animations and prototyping. We like everything! Right? So what happens is developers come into the picture and they give you sort of an idea of this. This is possible now; this is too much... Whatever. So that discussion is the most beautiful part of the entire process because there you have, like, a reality check on if it is possible or not or how can we best make it right.}''

\textit{(e) Brand Identity:} Designers often have to align with specific visual elements that reflect a brand's identity, such as color palettes, fonts, and styles. As P11 explained when choosing color palettes and other elements ``\textit{It's a matter of taste. And that's not always my taste. It sometimes is the taste of the marketing team or the client.}''

\textit{(f) Business:} Participants also noted that, especially in larger teams, the business strategy of the company often puts constraints on design. For example, P3 explained, ``\textit{Typically in a big company, you'll talk to product owners and managers and stakeholders to find out if it aligns with the business strategy, business values, all that stuff... in a smaller company you don't have as many mouths to feed so you get to kind of just go into it.}''

\subsubsection{Effects of content-based constraints on the design process.}
Participants had varied perceptions of having constraints during the ideation process. Some viewed the constraints as limitations, while others took a more positive approach, seeing constraints as helpful guides. 

\textbf{Negative Effects:}
Several participants disliked working under constraints, feeling limited and unable to ideate freely. For example, P8 noted, ``\textit{You want to do one thing, but then you have constraints that you need to come back to and think about them. There's a lot of back and forth between the ideation and making sure that you are following them.}'' Others, like P4, felt constraints led to suboptimal designs, in situations such as when corporate design systems restricted creativity: ``\textit{Maybe the project needs fancy design, but the design system limits us to some buttons and some styles that are defined before.}'' Participants also mentioned that working with constraints sometimes prevented them from thinking outside the box to make true innovations. P6 discussed this point: ``\textit{Maybe I haven't encountered a design guideline that is really different from my ideas. But you know I always have them in mind. So it affects my way of thinking while I'm working on a design.}''

\textbf{Positive Effects:}
Conversely, some participants, particularly those with more practical experience, saw constraints as beneficial. They felt constraints shape the ideation process, preventing distractions and speeding up ideation. By skipping decision-making on details like color palettes, they could focus on more important tasks. On this, P9 explained, ``\textit{Once you have a design system, creating new designs is quicker, letting you focus on where it's needed in that research and creation phase.}'' Many also likened constraints to ``\textit{Lego pieces,}'' with P3 saying they help structure designs while still allowing creativity: ``\textit{It's just a tool that helps you create your art piece.}'' Constraints also aid communication with developers and stakeholders, serving as a common language, as P11 said, ``\textit{They ensure developers know what's being proposed or if they already have one.}'' The design team can also use these constraints as a guiding ``north star'' to ease decision-making. As P15 explained, ``\textit{It's like an alignment within the team because a lot of designers focus on small details. So having those skylights helps us speak the same language when we're working together.}'' Moreover, some participants considered working under constraints as an essential aspect of design, which is parallel to the creative process, as P11 mentioned, ``\textit{I think the innovation and the ideation phase comes first. And then the prototyping and the design system come second. So in my process and in the process that I see other designers use, they're separate.}''

\subsubsection{Factors affecting the perception of content-based constraints:}
Several factors influenced designers' perception of working under constraints. One significant factor was \textbf{the explicitness of the constraints}. The content-based constraints could be either explicit or implicit, depending on how they were defined or communicated. \textit{Explicit constraints} are well-defined, clearly communicated, and easier for designers to navigate. For instance, P5 noted that ideation becomes straightforward when ``\textit{organizations prepare a creative brief beforehand, or the client already has something in mind and wants you to specifically follow that to stick to their brand identity.}'' Similarly, design systems can provide explicit structure, as P5 explained, ``\textit{Once the design system is sorted -- exact colors, sizes for headings, subheadings, and all that -- it becomes very clear.}'' \textit{Implicit constraints}, on the other hand, are less overt and often require designers' efforts to actively explore and extract them. For example, P8 elaborated that it is challenging to consider user characteristics when ``\textit{anticipating what it is the users are going to be looking for when they come to the website -- What would be the main categories or information that they would want to see...}'' P3 also emphasized the importance of understanding the implicit technical feasibility, considering a difficulty is to``\textit{validate that the solution we're offering is actually feasible};'' he continued, ``\textit{I used to be a dev, so I know what it takes to get something that is not possible.}''

Another factor was \textbf{the flexibility of the constraints}, which often depends on the openness of stakeholders for adjustments. A lack of flexibility added to the burden on the designers and created onerous processes. For example, P8 mentioned: ``\textit{There's a lot of back and forth between the ideation and making sure that you are following those constraints. It is a challenge.}'' On the other hand, the ability to adjust constraints allowed designers to push boundaries and innovate. For example, P13 described his experience: ``\textit{Sometimes I have arguments with the design system manager for adding a new component to make a design possible. And sometimes, after discussing the reasons for about 10 to 30 minutes, he accepts and adds it.}'' To attain this, however, required a certain level of experience from the designer to provide logical and well-reasoned arguments for deviating from a constraint. For example, P8, a seasoned designer who embraced constraints, mentioned, ``\textit{I haven't had anything that was so strict that has really affected [the project] too too much. ... Normally people are open to suggestions if you can explain why you made certain design decisions.}'' Participants mentioned the size of the company and team as another factor contributing to the openness and flexibility, as P3 highlighted: ``\textit{You just have to talk to more people in a big company.}''

Finally, \textbf{the availability of support from the team} played a crucial role in affecting the designers' perceptions of constraints. On this, P9 highlighted the importance of receiving support from the developers, stating, ``\textit{There's a collaboration with the development team that starts at the very beginning. We need the dev team's input on how we can potentially solve problems and what structure and tools we have in place to build new features.}'' P11 also shared: ``\textit{It's really good to chat with the engineering teams -- because the ideation phase should conform to something that's already in the system.}'' In addition to navigating technical constraints, support from the team also helped designers navigate business and industry-related constraints. For example, P3 noted its benefits: ``\textit{After a few iterations, it's the best to get feedback from product people -- sometimes you want to put sales on board, sometimes you want to include other departments.}''

\subsubsection{Designers' perception towards AI tools for design under constraints.}
\label{AI tools}
Given that our ultimate goal is to create an AI-powered tool to assist designers during the ideation process, we sought participants' perspectives on their current use of AI tools at this stage, the challenges they encountered, and the features they would like to see in ideal tools.

\textbf{Current Usage:} 
Participants used AI tools for ideation, layout structuring, and generating placeholder content. For example, P4 used AI for ``\textit{categorizing ideas and receiving suggestions,}'' P7 employed it for structuring page layouts, and some designers turned to AI tools for inspiration on colors. For example, P10 described using an extension in a web browser ``\textit{to see what specific colors people would use.}'' Similarly, many designers mentioned using AI to generate textual content, such as placeholder text. As P8 noted, ``\textit{You might use ChatGPT to get a little description that you don't want to have to write out yourself}'' A few participants discussed their use of AI tools to generate interactive prototypes as a foundation for further ideation and customization, as P11 shared: ``\textit{UIzard is able to create visual, clickable prototypes from a text prompt, which saves me hours or even days of work.}''

\textbf{Challenges and Desires:}
Designers preferred to play an active role when working with AI in shaping the concepts during ideation. For example, P5 said: ``\textit{You have to be very mindful when using AI... It can never replace real artists. That gap will always exist.}'' Participants expressed that they want AI to remain a source of inspiration rather than a replacement, with P5 adding: ``\textit{Not in a way that takes away the job of a UI/UX designer, but as an ongoing inspiration.}'' Participants wanted AI to generate diverse, relevant results while allowing designers to make final decisions. However, they emphasized that the results should not be random or generic and instead tailored to project goals and needs. On this, P3 expressed: ``\textit{It would be great to consider different brands, colors, and versions.}'' P10 highlighted industry-specific needs, saying: ``\textit{The medical industry should have different visuals and color tones rather than the fashion industry.}'' Participants also noted the need to modify generated results, as P1 stated: ``\textit{I want to be able to edit a specific part... What we have now is too general.}''

Moreover, participants identified a lack of familiarity and learnability with current AI-powered tools. Some tools offer editing features, but they are not intuitive for designers, as P10 noted: ``\textit{Making changes in UIzard isn't as intuitive as in Figma or Adobe XD.}'' Participants stressed that mastering these tools takes time and effort, as P9 suggested: ``\textit{There should be a help section or tutorial that shows the tool's full capabilities.}'' Nearly all participants noted that writing prompts for LLM-based tools is challenging and requires skill. On this, P9 said: ``\textit{You have to be very specific, and it may take some refining depending on the output.}'' Even after refining prompts, the desired outcome is not guaranteed, as P10 mentioned: ``\textit{The AI often doesn't understand what I want, even when I feel my prompt is complete.}'' Designers unfamiliar with the tool's capabilities struggled to create effective prompts. P12 said: ``\textit{I don't know how far the tool can help, so it's hard to write the right prompt.}'' Some participants suggested AI tools should improve input methods, with P5 proposing: ``\textit{It should ask questions and frame its own answers, rather than relying on one prompt like ChatGPT.}''

\subsection{Personas and Design Considerations}
Through our interview study, we identified three distinct designer groups based on their experience levels, each with unique perceptions of design constraints. For experienced designers, particularly those in leadership roles, constraints were valued for fostering consistency and efficiency within teams and across products. Junior designers, on the other hand, expressed frustration with constraints, perceiving them as restrictive to their creative exploration. Conversely, recently graduated design students viewed constraints positively as they provided structure and guidance to compensate for the lack of experience. To better address the needs of these groups, we developed three personas to represent each group, respectively. These personas were created by revisiting interview data and grouping codes from related participants to extract relevant insights. They were named with alliteration: Eric the experienced designer, Julia the junior designer, and Sarah the student entering the job market. We briefly introduce the personas here; their detailed information can be found in Appendix~\ref{sec:app_personas}.

\begin{enumerate}[leftmargin=18pt]
    \item \textbf{Eric} is a seasoned UI/UX designer with 12 years of experience, leading a team at a tech company. He values structured creativity and uses constraints to promote consistency and efficiency in his team's work. His challenges include balancing innovation while maintaining a cohesive design language and effectively communicating design changes with stakeholders.
    
    \item \textbf{Julia} is a freelance UI/UX designer with four years of experience, seeking creative freedom while balancing client demands. She struggles with managing client expectations, working under tight deadlines, and ensuring technical feasibility.

    \item \textbf{Sarah} is a design student preparing to enter the workforce, who is eager to learn and apply her theoretical knowledge to practical projects. Sarah faces challenges in selecting cohesive design elements and lacks access to professional resources.
\end{enumerate}

\input{table_designideas}

After persona creation, we first conducted brainstorming sessions to come up with concrete design ideas and actions that could address the main goals and frustrations of each persona. Then we performed an affinity diagramming exercise to group these design ideas and actions into higher-level design considerations (DCs) that can inform UIDEC and other tools for design ideation under constraints. Overall, the primary focus of our tool is on \textbf{supporting UI/UX designers in creative inspiration}, rather than serving as a full-fledged prototyping solution. With this scope in mind, we consciously avoided considering aspects that aimed to support prototyping or developer handoff. Table~\ref{tab:design_ideas} summarizes our design considerations, along with how each is linked to the personas to address their unique needs and challenges. We briefly describe these considerations below.

\textbf{DC1: Integrating in designers' early-stage processes to maximize creative exploration.}
Eric requires tools to integrate past work and foster team collaboration, while Julia values compatibility with external design tools for her freelance projects. Ideation tools should thus support the collaborative process and integrate seamlessly into the designer's workflow.

\textbf{DC2: Providing scaffolding for creative inspiration to minimize uncertainty and confusion.}
Sarah, as a novice designer, needs guidance when using new tools, exploring design ideas, and staying updated on industry trends. To address these needs, tools for ideation under constraints should provide guidance and inform designers of domain-specific common practices and patterns.

\textbf{DC3: Facilitating flexibility in defining constraints}
Eric and Julia both need flexibility when defining design constraints -- Eric for satisfying diversity across projects, and Julia for adapting to diverse clients' needs. So, enhancing flexibility and freedom during constraint specification is important.

\textbf{DC4: Allowing exploration of design alternatives through iterative modifications}
Sarah needed features to track her inspiration and ideation process, while Julia wanted to explore design variations for different and evolving client needs. Thus, it is critical for tools to support design exploration during the iterative process of creative design.

\textbf{DC5: Facilitating the organization of ideas based on projects and preferences}
Sarah needs features to help develop her design style, while Julia requires tools to manage multiple client projects. Thus, ideation tools should allow designers to organize ideas in flexible ways to support their work.

%% file: table_designideas.tex
\begin{table*}[t]
\centering
\caption{Summary of our design ideas and actions, mapped to persona's requirements and design considerations}
\label{tab:design_ideas}
\begin{tabular}{p{.2\textwidth}p{.35\textwidth}p{.35\textwidth}}
\toprule
\textbf{Design Considerations} &
  \textbf{Persona's Requirements} &
  \textbf{Potential Design Ideas and Actions} \\
  
\midrule
\multirow{1}{.2\textwidth}{DC1. Integrating in designers' early-stage processes to maximize creative exploration.} &
  [Eric] Facilitating collaboration with team members and stakeholders. &
  \begin{minipage}[t]{\linewidth}
  \begin{itemize}[leftmargin=*]
    \item[-] Exporting the designs and constraints to share with team members and stakeholders.
  \end{itemize}
  \end{minipage}\\
 &
  [Eric] Integrating previous work to improve consistency and continuity. &
  \begin{minipage}[t]{\linewidth}
  \begin{itemize}[leftmargin=*]
    \item[-] Ability to upload previous work to guide design generation. [Not implemented]
  \end{itemize}
  \end{minipage}\\
 &
  [Julia] Being compatible with other design tools. &
  \begin{minipage}[t]{\linewidth}
  \begin{itemize}[leftmargin=*]
    \item[-] Having different formats for exporting the designs so that designers can import them into other design tools. [Not implemented]
  \end{itemize}
  \end{minipage}
  \vspace{2pt}\\

\midrule
\multirow{1}{.2\textwidth}{DC2. Providing scaffolding for creative inspiration to minimize uncertainty and  confusion.} &
  [Sarah] Showing the capabilities of the tool to the users so that they know what to expect from the tool. &
  \begin{minipage}[t]{\linewidth}
  \begin{itemize}[leftmargin=*]
    \item[-] Showing some generated designs and the related constraints when the tool is loaded the first time.
  \end{itemize}
  \end{minipage}\\
 &
  [Sarah] Keeping UI/UX designers updated with the latest design trends. &
  \begin{minipage}[t]{\linewidth}
  \begin{itemize}[leftmargin=*]
    \item[-] Providing suggestions on design themes consisting of fonts, colors, and component styles, based on what is trending.
  \end{itemize}
  \end{minipage}\\
 &
  [Sarah] Reducing confusion while choosing design constraints. &
  \begin{minipage}[t]{\linewidth}
  \begin{itemize}[leftmargin=*]
    \item[-] Showing hints or suggestions when selecting different constraints. (E.g., Designers mostly use these color palettes when designing health apps.) [Not implemented]
  \end{itemize}
  \end{minipage}
  \vspace{2pt}\\
  
\midrule
\multirow{1}{.2\textwidth}{DC3. Facilitating flexibility in defining constraints.} &
  [Eric] Providing flexibility in specifying the functionality of the design. &
  \begin{minipage}[t]{\linewidth}
  \begin{itemize}[leftmargin=*]
    \item[-] Providing options for defining functionality and screen types.
  \end{itemize}
  \end{minipage}\\
 &
  [Eric] Allowing different ways to effectively describe the desired design constraints. &
  \begin{minipage}[t]{\linewidth}
  \begin{itemize}[leftmargin=*]
    \item[-] Having multiple options to choose from regarding the design constraints and providing text inputs to describe more details.
  \end{itemize}
  \end{minipage}\\
 &
  [Julia] Making a design constraint constant during the generation process. &
  \begin{minipage}[t]{\linewidth}
  \begin{itemize}[leftmargin=*]
    \item[-] Having a ``Lock'' option on the constraints. When locked, all the generated examples should observe that constraint.
  \end{itemize}
  \end{minipage}
  \vspace{2pt}\\

\midrule
\multirow{1}{.2\textwidth}{DC4. Allowing exploration of design alternatives through iterative modifications.} &
  [Sarah] Being able to modify different aspects of the generated results. &
  \begin{minipage}[t]{\linewidth}
  \begin{itemize}[leftmargin=*]
    \item[-] Providing a dialog box to allow users to adjust options and provide inputs to change in the design.
  \end{itemize}
  \end{minipage}\\
 &
  [Julia] Being able to show different variations of the generated design. &
  \begin{minipage}[t]{\linewidth}
  \begin{itemize}[leftmargin=*]
    \item[-] Having a ``Regenerate'' feature, while keeping the previous versions when editing a generated design.
  \end{itemize}
  \end{minipage}
  \vspace{2pt}\\

\midrule
\multirow{1}{.2\textwidth}{DC5. Facilitating the organization of ideas based on projects and preferences.} &
  [Sarah] Being able to save their favorite designs. &
  \begin{minipage}[t]{\linewidth}
  \begin{itemize}[leftmargin=*]
    \item[-] Having a ``Like'' feature in generated examples and saving them in folders to create personal mood boards.
  \end{itemize}
  \end{minipage}\\
 &
  [Julia] Organizing user's different work in multiple projects to facilitate easy access. &
  \begin{minipage}[t]{\linewidth}
  \begin{itemize}[leftmargin=*]
    \item[-] Having a ``Canvas Collection'' feature to manage projects.
  \end{itemize}
  \end{minipage}
  \vspace{2pt}\\
  \bottomrule
\end{tabular}
\end{table*}

%% file: s_systemdesign.tex
\section{UIDEC Design and Implementation}

\subsection{Design and Implementation Process}
Following the creation of personas and the design considerations, the design ideation phase was initiated by two of the authors, generating tentative design concepts that aligned with the identified user needs. These initial ideas were then presented and discussed in weekly team meetings with all the authors to refine the concepts and ensure that the designs aligned with the project goals.
Next, a site map and user flow were developed, providing a comprehensive overview of the tool's structure and user journey. A list of concrete feature design ideas and actions was then created, with each feature carefully identified and mapped to specific personas to ensure the tool would cater effectively to different user scenarios (see Table~\ref{tab:design_ideas}). Sketches and prototypes of each feature were then produced, reviewed, and iterated upon through team discussions. Along this process, an interactive Figma prototype was created and a working system was implemented by two of the authors. Throughout the development phase, the tool was tested iteratively by the entire team, ensuring that it met the design objectives.

\subsection{Interaction Design}
\label{sec:features}
UIDEC has a minimalist interface. Upon first visiting the tool, users are presented with a curated set of examples and detailed specifications, demonstrating the platform's capabilities and guiding new users through its features [DC2]. Once logged in, the interface transforms into a fully functional workspace. Below, we introduce the interaction design of UIDEC through three workflows.

\label{sec:design_workflow1}
\begin{figure*}[t]
    \centering
    \includegraphics[width=0.93\textwidth]{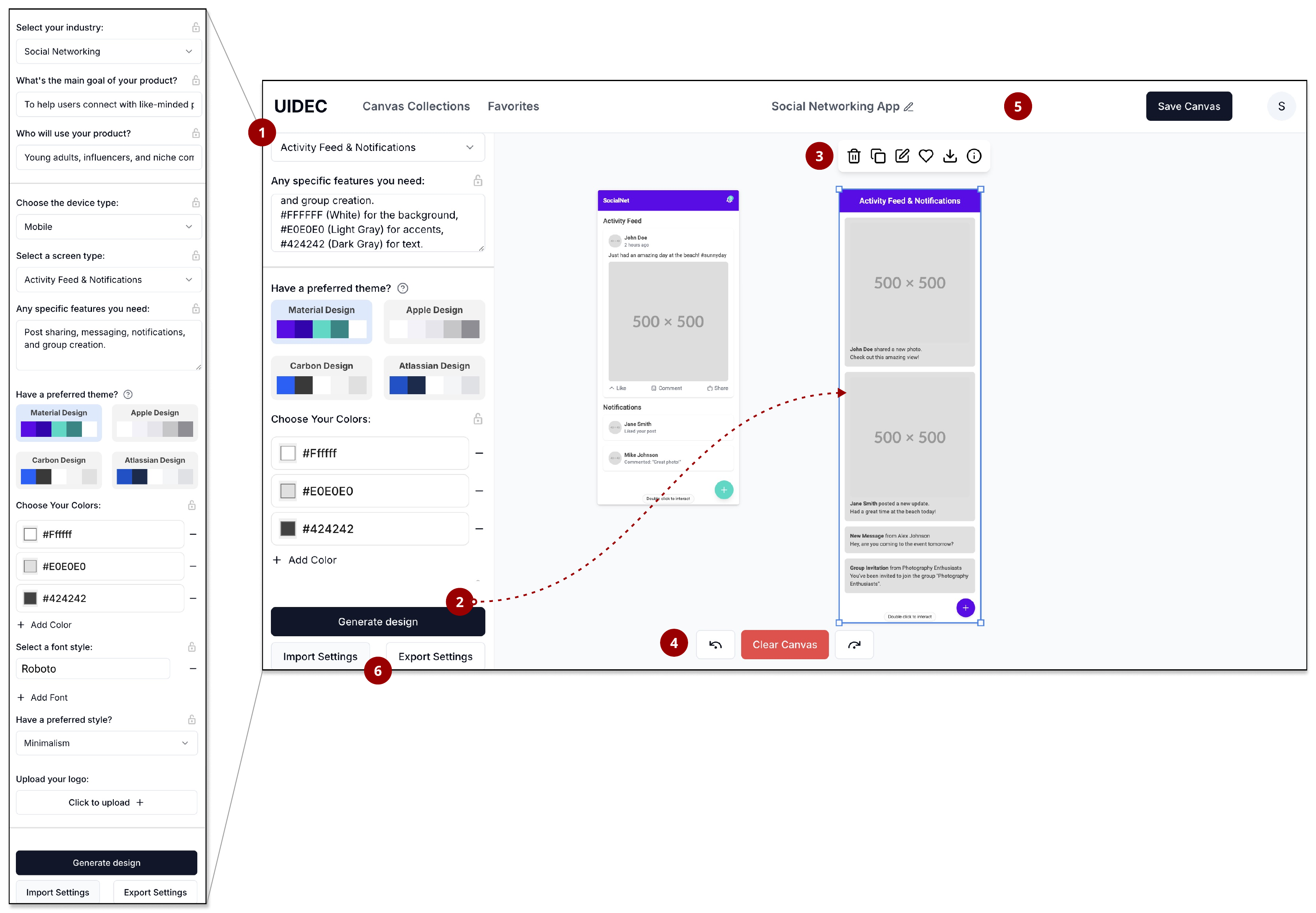}
    \caption{UIDEC interaction workflow 1: generating the design examples}
    \label{fig:design_workflow1}
    \Description{This figure shows the main interface of UIDEC, divided into several key areas. On the left side is a panel with various input fields for specifying design constraints, labeled with (1). These inputs include dropdowns for industry, device type, and screen type, as well as text fields for product goals and target audience. The panel also includes options for choosing colors, fonts, and design themes. The central area of the interface shows the main canvas, labeled (2), where a generated design is displayed. This design appears to be a mobile app interface for a social networking application, showing an activity feed and notifications. Above the generated design is a toolbar, labeled (3), which includes icons for various actions such as deleting, duplicating, editing, saving, and downloading the design. At the bottom of the canvas are control buttons labeled (4), including options to undo, redo, and clear the canvas. The top of the interface shows a navigation bar, with options for Canvas Collections and Favorites. On the far right of this bar is a "Save Canvas" button, labeled (5). Additionally, at the bottom left of the figure are buttons for importing and exporting settings, labeled (6). In sum, this figure illustrates the process of inputting design constraints and generating a design example in UIDEC.}
\end{figure*}

\subsubsection{Workflow 1: Generating the Design Examples}
The main interface of UIDEC features a side panel, a main canvas, and a top navigation bar (see Figure~\ref{fig:design_workflow1}). The side panel ((\textcircled{1} in Figure~\ref{fig:design_workflow1}) includes various inputs corresponding to the constraints designers face in the ideation process, as described in section~\ref{sec:interview_constraints}. Using these inputs, users can select the industry, specify the goal of their product, define their target users, and choose the device and screen type [DC3]. They can also describe specific features they need and select a preferred design theme [DC2]. Additionally, users can select up to five colors and three fonts, choose their preferred design style, and upload their logo. Users can lock specific constraints to ensure they remain fixed throughout the entire design idea generation process for consistency [DC3]. Once the inputs are provided, users can click the \textit{Generate design} button, and a design (\textcircled{2} in Figure~\ref{fig:design_workflow1}) is produced on the canvas.

Upon selecting the generated design, a toolbar (\textcircled{3} in Figure~\ref{fig:design_workflow1}) appears with six buttons, respectively for removing the design, duplicating the design, editing the design, saving the design to a Favorites folder, downloading the HTML file of the design, and displaying the design specifications [DC1, DC4, DC5]. Users can also move and resize the designs freely on the canvas [DC5]. The canvas also includes three control buttons: \textit{Undo}, \textit{Redo}, and \textit{Clear Canvas} (\textcircled{4} in Figure~\ref{fig:design_workflow1}) [DC4]. Additionally, the UIDEC allows users to rename and save their canvas (\textcircled{5} in Figure~\ref{fig:design_workflow1}) from the top navigation bar [DC5]. Canvas organization is further described in Section~\ref{sec:design_workflow3}.

Users can generate additional designs by pressing the \textit{Generate Design} button as many times as they wish, with each click producing a new design next to the previously generated design. The canvas automatically zooms in to display the newly generated design. Below the \textit{Generate Design} button in the left side panel, there are two additional options: \textit{Import Settings} (to load constraints and settings from a JSON file) and \textit{Export Settings} (to save the current constraints and settings as a JSON file) (\textcircled{6} in Figure~\ref{fig:design_workflow1}) [DC1].

\label{sec:design_workflow2}
\begin{figure*}[t]
    \centering
    \includegraphics[width=0.6\textwidth]{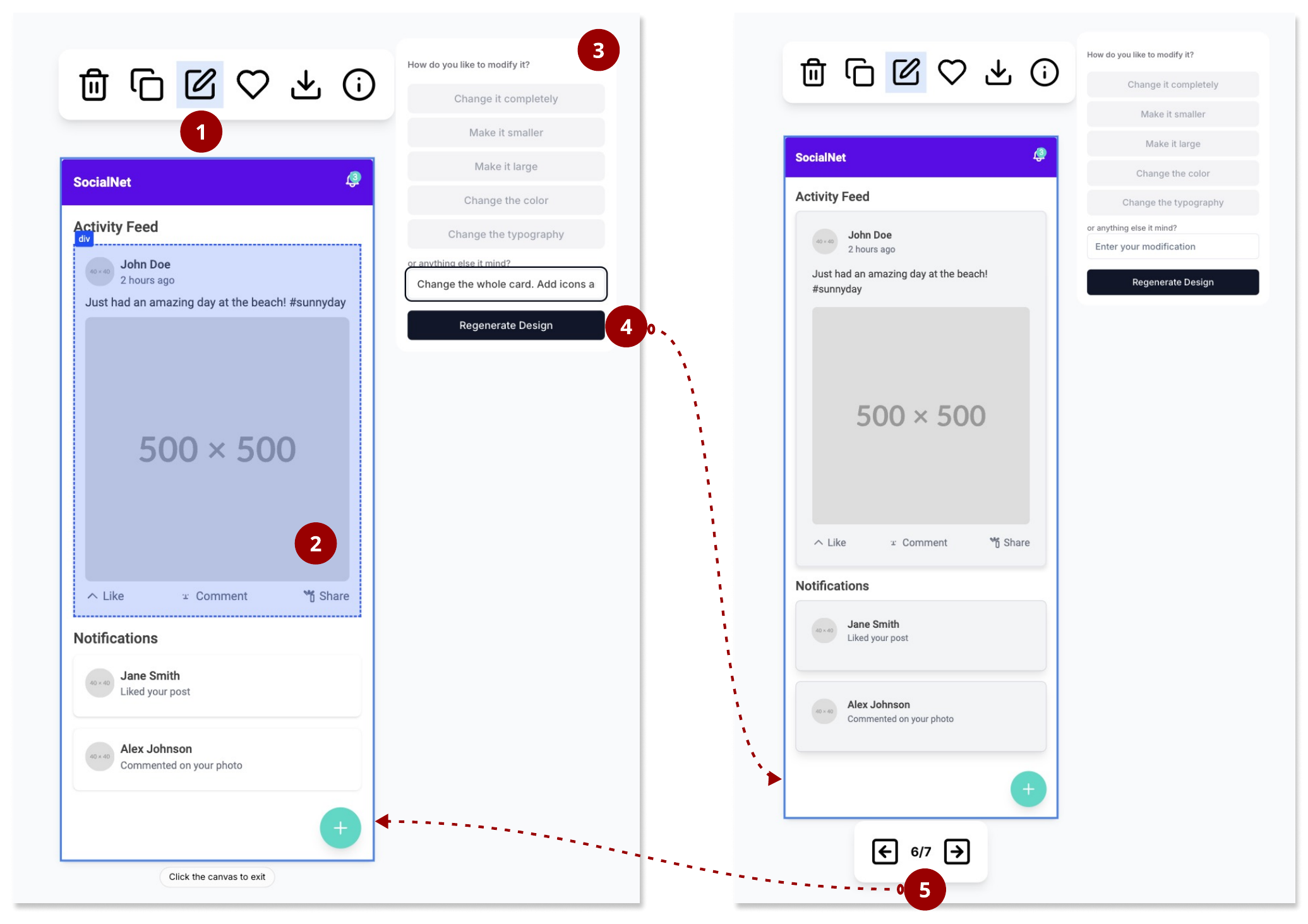}
    \caption{UIDEC interaction workflow 2: editing the design examples}
    \label{fig:design_workflow2}
    \Description{This figure demonstrates the process of editing a generated design in UIDEC. The interface is similar to Figure 1, but with additional elements related to the editing process. In the top left corner, labeled (1), is an "Edit" button in the toolbar of a generated design. The central area shows a generated design with a specific element highlighted, labeled (2), indicating that the user has selected this part for editing. To the right of the design is a "Modification Box", labeled (3). This box includes options for adjusting the selected element, such as resizing, changing colors, and modifying typography. There's also a text box for entering detailed modification requests. Below the Modification Box is a "Regenerate Design" button, labeled (4), which the user can click to apply their requested changes. At the bottom of the figure, labeled (5), is a navigation panel with left and right arrows and numbered indicators. This allows users to browse through different versions of their edited designs. In sum, this figure illustrates UIDEC's capabilities for iterative design modification and version control.}
\end{figure*}

\subsubsection{Workflow 2: Editing the Design Examples}
Once a design is generated, users can make edits by using the \textit{Edit} button (\textcircled{1} in Figure~\ref{fig:design_workflow2}) located in the toolbar [DC4]. After clicking the \textit{Edit} button, users must select the specific part (\textcircled{2} in Figure~\ref{fig:design_workflow2}) of the design they wish to modify. Upon selection, a \textit{Modification Box} (\textcircled{3} in Figure~\ref{fig:design_workflow2}) appears next to the design, offering various options such as resizing the component (smaller or larger), altering the color scheme or typography, as well as a text box allowing users to write detailed modification requests [DC4]. Once the desired changes are set, users can click the \textit{Regenerate Design} button (\textcircled{4} in Figure~\ref{fig:design_workflow2}) to generate a new version of the design with the applied modifications. Below the generated design, a navigation panel (\textcircled{5} in Figure~\ref{fig:design_workflow2}), with left and right arrows and numbered indicators, enables users to browse through earlier versions of their edited designs [DC4].

\label{sec:design_workflow3}
\begin{figure*}[t]
    \centering
    \includegraphics[width=0.95\textwidth]{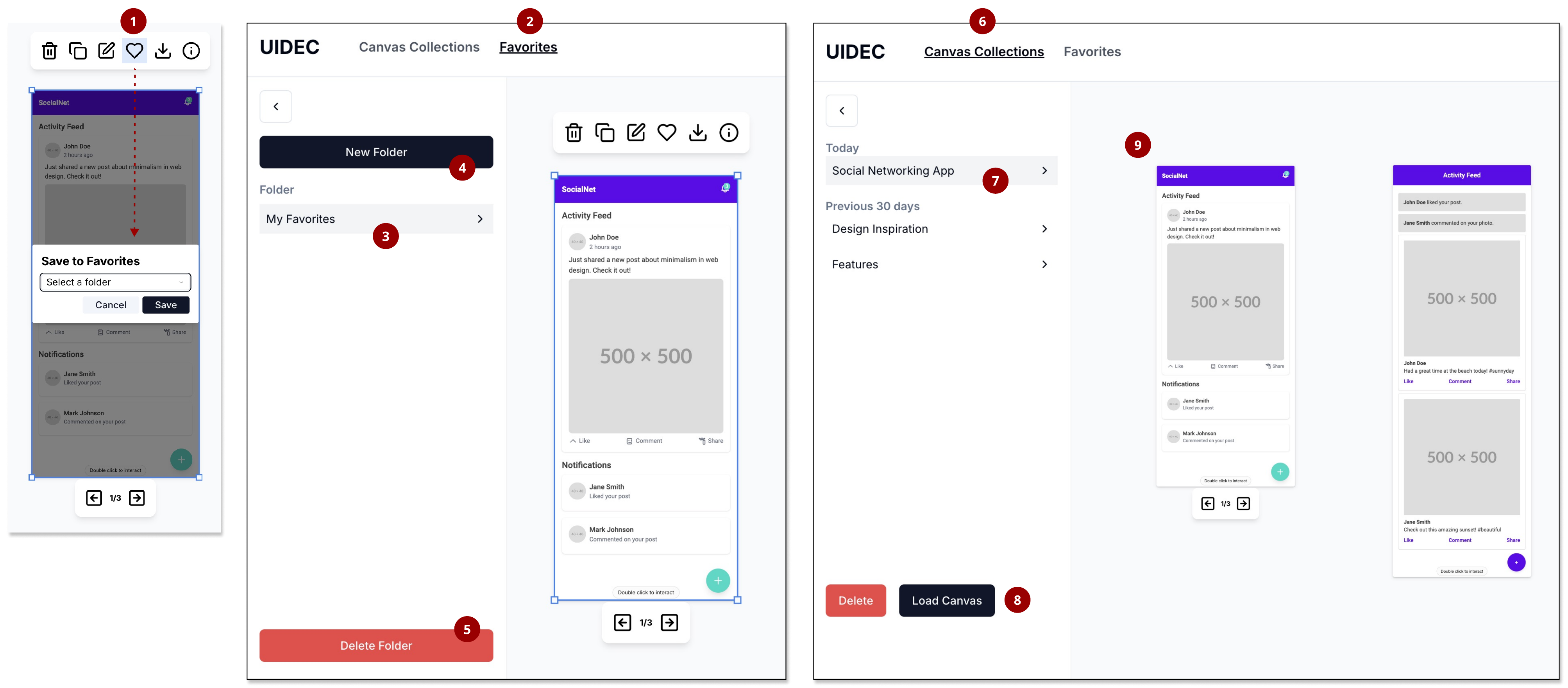}
    \caption{UIDEC interaction workflow 3: organizing the design examples}
    \label{fig:design_workflow3}
    \Description{This figure showcases the design organization features of UIDEC. It's divided into several key areas, each labeled with a number. Label (1) points to a "Save to Favorites" popup, which appears when a user wants to save a design. This popup allows users to select a folder for saving their design. At the top of the interface is a navigation bar. Label (2) indicates the "Favorites" button in this bar. On the left side of the interface is a panel for folder selection, labeled (3). Below this are options to create a new folder (4) and delete existing folders (5). The top navigation bar also includes a "Canvas Collections" option, labeled (6). When selected, this displays saved canvases in the left panel, as indicated by label (7). When a canvas is selected, two options appear: "Load Canvas" and "Delete", labeled (8). On the right side of the interface, labeled (9), is a preview of the selected canvas. In sum, this figure demonstrates UIDEC's features for organizing and managing multiple designs and projects.}
\end{figure*}

\subsubsection{Workflow 3: Organizing the Design Examples}
Users can save any generated design they like to a \textit{Favorites} folder [DC5]. To do this, the user clicks the \textit{Save} button in the design's toolbar and a \textit{Save to Favorites} popup (\textcircled{1} in Figure~\ref{fig:design_workflow3}) will appear, prompting the user to select a folder and confirm by clicking the \textit{Save} button on the popup. Users can access their saved designs by clicking on \textit{Favorites} (\textcircled{2} in Figure~\ref{fig:design_workflow3}) in the top navigation bar. There, users can select the desired folder (\textcircled{3} in Figure~\ref{fig:design_workflow3}) from the left side panel, and the canvas on the right will display all the designs saved in that folder. Users can also create new folders (\textcircled{4} in Figure~\ref{fig:design_workflow3}) or delete existing folders (\textcircled{5} in Figure~\ref{fig:design_workflow3}). This feature helps users build mood boards, allowing them to categorize and store design inspirations based on their specific needs [DC5].

Additionally, saved canvases (see Section~\ref{sec:design_workflow1}) can be accessed via the \textit{Canvas Collections} present in the top navigation bar (\textcircled{6} in Figure~\ref{fig:design_workflow3}) [DC5]. The saved canvases are displayed in the left side panel. Upon selecting a canvas (\textcircled{7} in Figure~\ref{fig:design_workflow3}), two options, \textit{Load Canvas} and \textit{Delete}, appear in the left panel (\textcircled{8} in Figure~\ref{fig:design_workflow3}), while a preview of the selected canvas is displayed on the right (\textcircled{9} in Figure~\ref{fig:design_workflow3}). By clicking \textit{Load Canvas}, users can open the desired canvas with all the original settings and designs for further ideation.

\subsection{Implementation}
\subsubsection{System Architecture}
Our system leverages modern web technologies and a Large Language Model (LLM) to generate UI designs based on user-specified constraints. The application is built using Next.js, a React framework that provides both frontend and backend capabilities \cite{nextjs}. For authentication, database management, and storage, we integrate Pocketbase, an open-source backend solution \cite{pocketbase}. The core of our design generation process relies on the OpenAI API, specifically utilizing the GPT-4o model for its advanced language understanding, multi-modal capabilities, and code generation capabilities \cite{openai}.

The system follows a monolithic architecture, where frontend and backend are tightly integrated within the Next.js framework, simplifying deployment and management by using a single codebase that handles UI, backend logic, and database interaction. The frontend, implemented using React components, communicates with the backend through Next.js API routes. User inputs are processed in the backend, where appropriate prompts are constructed and sent to the OpenAI API to generate an HTML page that satisfies the constraints, with added variations each time the page is generated (see Section~\ref{sec:impl_prompt}). The OpenAI API outputs are then parsed and the generated HTML designs are returned to the frontend for rendering on the canvas.

To ensure efficient rendering and manipulation of multiple designs simultaneously, we employ the HTML canvas API \cite{html_canvas}. This approach allows for smooth user interaction with generated designs, enabling features such as zooming, panning, and selecting specific elements. The canvas implementation includes custom drawing and interaction functions, allowing the user to freely resize and update the generated designs.

Constraint adherence is mostly achieved through prompt construction (see Section~\ref{sec:impl_prompt}). Additionally, the ``Device'' constraint (mobile, tablet, desktop) is reinforced by setting the size and aspect ratio of the viewport that renders the generated HTML page. However, users can resize the viewport to explore how the UI will appear and function on different screens.

\subsubsection{Prompt Construction}
\label{sec:impl_prompt}
The prompt construction process is crucial for generating UI designs that align with user specifications while providing diversity and variation. Our prompt for generating UI designs includes two main parts: the system prompt and the user prompt. The system prompt (provided in Appendix \ref{sec:app_system_prompt}) sets the context for the AI, positioning it as an experienced web designer and developer. Table~\ref{tab:system-prompt-justification} justifies the structure of the system prompt by outlining each component and its rationale.

\begin{table*}[t]
\centering
\caption{System prompt structure and justifications for its components}
\label{tab:system-prompt-justification}
\begin{tabular}{p{0.3\textwidth}p{0.65\textwidth}}
\toprule
\textbf{Prompt Component} & \textbf{Justification} \\
\midrule
\textbf{Role Definition} & The AI is positioned as an ``exceptional web designer and developer with millennia of experience'' to ensure the model takes on the role of an expert capable of creating high-quality and modern website prototypes. \vspace{4pt}\\

\textbf{Expertise Scope} & By emphasizing that the AI's knowledge spans ``countless design trends, technologies, and best practices,'' the model is encouraged to leverage its knowledge of both contemporary and historical web development techniques, ensuring that outputs are well-rounded and contextually appropriate. \vspace{4pt}\\

\textbf{Task Definition} & The AI is explicitly tasked with ``transforming specific requirements into visually stunning and functional websites,'' focusing its efforts on producing appealing and functional outputs that adhere to specific user inputs and constraints. \vspace{4pt}\\

\textbf{Understanding User Specifications} & The AI is provided clear instructions to carefully analyze specifications such as industry context, colors, fonts, devices, design themes, screen types, target audience, and product purpose. This ensures that the model generates tailored designs that meet the user's precise requirements. \vspace{4pt} \\

\textbf{Interpreting Design References} & When given reference UI screens, the AI is instructed to focus on layout and structure while disregarding non-relevant elements such as colors and fonts, unless specified otherwise. This helps ensure consistency in following user-defined specifications while leveraging external design references for guidance. \vspace{4pt} \\

\textbf{Prototype Creation Guidelines} & Detailed prototype creation instructions—using Tailwind CSS, custom CSS/JS, Google Fonts, and placeholder images—are provided to ensure that the AI follows best practices in producing a fully responsive and interactive website prototype. \vspace{4pt} \\

\textbf{Result Presentation} & The AI is instructed to deliver its response as a single HTML file with an interactive prototype, ensuring a cohesive and usable design output that can be easily reviewed and integrated by the user. \\
\bottomrule
\end{tabular}
\end{table*}

The user prompt is constructed dynamically based on user inputs. It combines the following elements:

\begin{itemize}
    \item The \textit{base prompt} (see Appendix~\ref{sec:app_user_prompt:base}) sets the context for the AI, framing the task as a real-world design request from a product manager. It also clearly specifies the expected format of the response.
    \item The \textit{user specifications prompt} (see Appendix~\ref{sec:app_user_prompt:specs}) includes the user's design constraints specified through the UI, such as industry, screen type, color palette, and fonts.
    \item A \textit{design theme expansion} will be appended when the design theme is selected in user specifications. It includes detailed specifications for colors, typography, and component styles for the selected theme. An example of the Material Design theme is included in Appendix~\ref{sec:app_user_prompt:theme}.
    \item An \textit{reference UI screen prompt} (see Appendix~\ref{sec:app_user_prompt:UI}) instructs the AI to use the provided UI screen as a structural reference while disregarding specific design elements that may conflict with the user's specifications. This prompt is added to enhance the variation and the diversity of the generated images. The details of the UI screen dataset and the selection process are described below.
\end{itemize}

To improve layout diversity in the generated UI designs, we incorporated \textit{reference UI screens} as structural guidance for design generation. A reference screen was randomly selected according to the user constraints from a dataset derived from the Mobbin platform \cite{mobbin}. We manually curated this dataset by collecting UI screens from Mobbin along three main constraints: industry categories, screen types, and device types; these constraints were selected to organize the dataset because they often indicate distinctive page layouts and structural styles. For each combination of possible values of the industries and screen types (as options provided to the user in UIDEC), we browsed the Mobbin platform for corresponding apps and websites and selected up to 50 unique UI screens across Mobile and Desktop device types. Through this process, we collected a dataset that comprises 14640 UI screens from iOS, Android, and web applications. For each combination of user-specified constraints, the database contains multiple images from different applications, thus offering a rich source of diverse design patterns and layouts. After the dataset was collected, we then converted all images to grayscale, aiming to reduce the impact of color on the model's interpretation of the design while preserving the essential layout and structural information. During the prompt construction process, we first query the dataset for records that match both the user-specified industry category and screen type. If matching records are found, we randomly select one image from the set of images in the matching records. If no matching records are found, no image is selected, and the final prompt will not include an image reference.

UIDEC also supports \textit{editing previously generated designs}. After the user has selected a section they want to edit, the tool allows them to specify the edit they want or select from a list of common editing commands (see Section~\ref{sec:design_workflow2}). UIDEC then constructs the edit prompt based on the editing task and content of the previous design, along with the system prompt described above. The detailed edit prompt is shown in Appendix~\ref{sec:edit_design_prompt:edit}.

By carefully constructing the prompt with user constraints and providing relevant information and reference images, we ensure that the LLM generates HTML code for UI designs that closely align with user specifications while maintaining creative freedom in areas not explicitly constrained. This approach allows for generating unique and tailored UI designs that meet specific project requirements while leveraging the power of advanced language models and previous design examples and principles.

\subsection{Constraint Adherence Evaluation}

\begin{table}[t]
\centering
\caption{Summary of design constraint sets for evaluating adherence to color schemes, fonts, device types, and logo inclusion}
\label{tab:design-sets-rq1}
\resizebox{\columnwidth}{!}{
\begin{tabular}[\columnwidth]{cllcc}
\toprule
\textbf{ID} & \textbf{Colors} & \textbf{Fonts} & \textbf{Device} & \textbf{Logo} \\
\midrule
1 & \fcolorbox[HTML]{BBBBBB}{2C3E50}{\rule{0pt}{3pt}\rule{3pt}{0pt}} \fcolorbox[HTML]{BBBBBB}{18BC9C}{\rule{0pt}{3pt}\rule{3pt}{0pt}} \fcolorbox[HTML]{BBBBBB}{ECF0F1}{\rule{0pt}{3pt}\rule{3pt}{0pt}} & Orelega One, Pacifico, Montserrat & Desktop & Yes \\
2 & \fcolorbox[HTML]{BBBBBB}{2196F3}{\rule{0pt}{3pt}\rule{3pt}{0pt}} \fcolorbox[HTML]{BBBBBB}{FFFFFF}{\rule{0pt}{3pt}\rule{3pt}{0pt}} \fcolorbox[HTML]{BBBBBB}{FFC107}{\rule{0pt}{3pt}\rule{3pt}{0pt}} & Merriweather, Philosopher, Platypi & Tablet & Yes \\
3 & \fcolorbox[HTML]{BBBBBB}{4CAF50}{\rule{0pt}{3pt}\rule{3pt}{0pt}} \fcolorbox[HTML]{BBBBBB}{FFFFFF}{\rule{0pt}{3pt}\rule{3pt}{0pt}} \fcolorbox[HTML]{BBBBBB}{FF5722}{\rule{0pt}{3pt}\rule{3pt}{0pt}} & Lato, Prompt, Quando & Mobile & Yes \\
4 & \fcolorbox[HTML]{BBBBBB}{212121}{\rule{0pt}{3pt}\rule{3pt}{0pt}} \fcolorbox[HTML]{BBBBBB}{FFEB3B}{\rule{0pt}{3pt}\rule{3pt}{0pt}} \fcolorbox[HTML]{BBBBBB}{E91E63}{\rule{0pt}{3pt}\rule{3pt}{0pt}} & Montserrat, Revalia, Playfair & Desktop & Yes \\
5 & \fcolorbox[HTML]{BBBBBB}{000000}{\rule{0pt}{3pt}\rule{3pt}{0pt}} \fcolorbox[HTML]{BBBBBB}{FF0000}{\rule{0pt}{3pt}\rule{3pt}{0pt}} \fcolorbox[HTML]{BBBBBB}{FFFFFF}{\rule{0pt}{3pt}\rule{3pt}{0pt}} & Roboto, Rubik, Silkscreen & Mobile & Yes \\
\bottomrule
\end{tabular}
}
\end{table}

We conducted a preliminary evaluation to assess the adherence of our generated UI designs to specified constraints. For constraints like \textit{color schemes}, \textit{fonts}, \textit{device types}, and \textit{logo inclusion}, we can directly inspect the output HTML code to evaluate whether they are observed. For this purpose, we created five diverse sets of design briefs (summarized in Table~\ref{tab:design-sets-rq1}), each incorporating a unique combination of constraints. For each set, our approach generated five design variations, resulting in a total of 25 generated designs. We then measured how well each design adhered to the specified objective constraints, using the following formula:
\begin{equation*}
\begin{aligned}
&\text{Adherence}(c) = \\
&\hspace{12pt}\frac{\text{Number of correctly implemented instances of } c}{\text{Total number of instances where} c \text{ should be applied}} \times 100\%
\end{aligned}
\end{equation*}

The results show strong adherence to objective constraints across all sets, with perfect adherence to color schemes, device types, and logo inclusion. Font adherence shows some variation among the five sets, ranging from perfect adherence to 40\%, with an average of 74.7\% across all sets.

\begin{figure*}[!ht]
    \centering
    \includegraphics[width=0.99\textwidth]{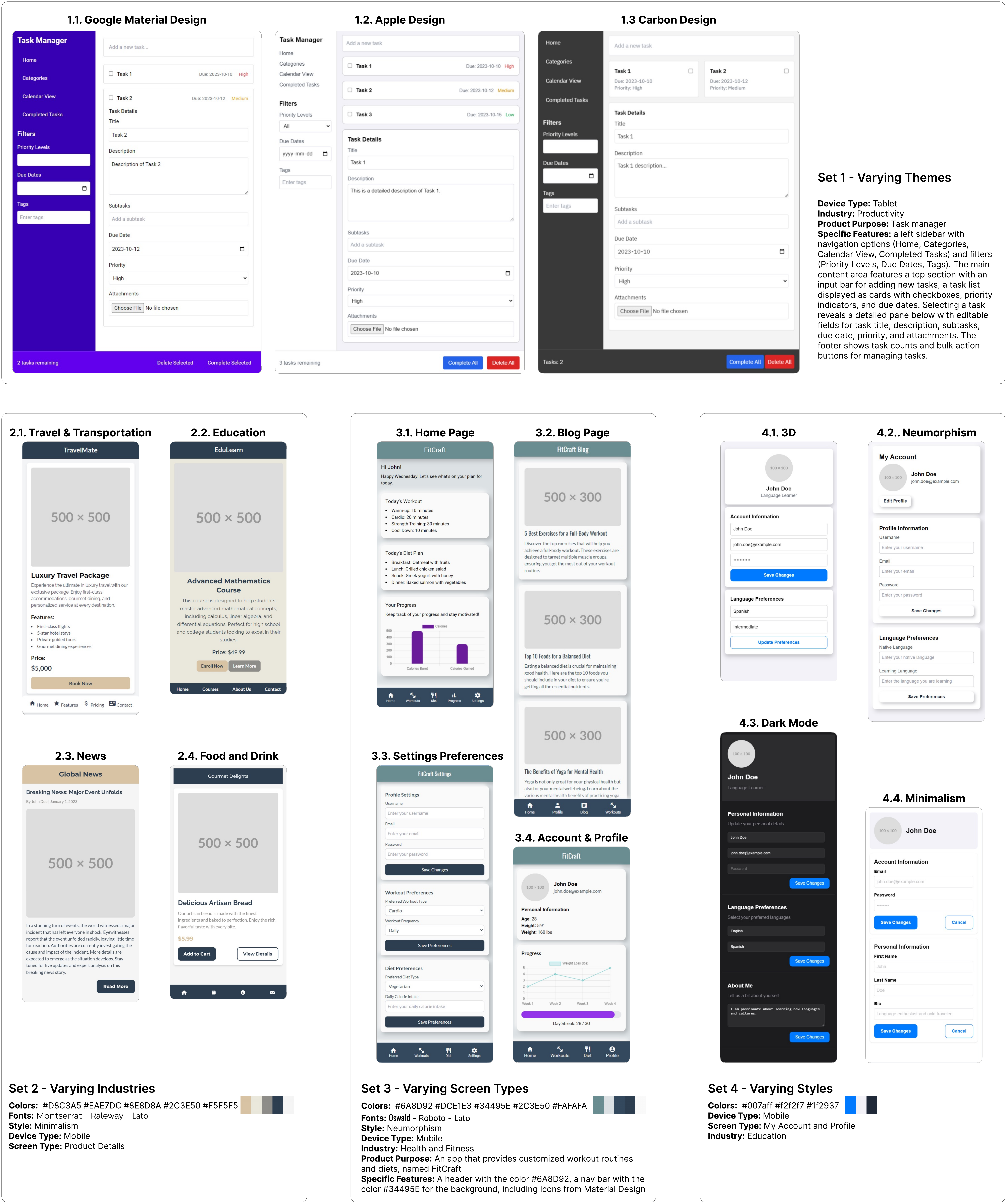}
    \caption{Examples of the generated designs when varying a specific constraint while keeping others constant}
    \label{fig:variations}
    \Description{This figure is divided into four sets, each demonstrating how UIDEC responds to changes in a specific constraint while keeping others constant. Set 1 shows designs with varying themes. It includes three designs labeled 1.1 (Google Material Design), 1.2 (Apple Design), and 1.3 (Carbon Design). Each design shows a task manager interface for a productivity app on a tablet device. Set 2 demonstrates designs for different industries. It includes four designs labeled 2.1 (Travel & Transportation), 2.2 (Education), 2.3 (News), and 2.4 (Food and Drink). These designs show product detail pages for mobile devices, all following a minimalist style. Set 3 presents designs for various screen types within the Health and Fitness industry. It includes four designs labeled 3.1 (Home Page), 3.2 (Blog Page), 3.3 (Settings Preferences), and 3.4 (Account & Profile). These are all mobile designs following a neumorphic style. Set 4 illustrates designs with different styles for an educational app. It includes four designs labeled 4.1 (3D), 4.2 (Neumorphism), 4.3 (Dark Mode), and 4.4 (Minimalism). These show account and profile pages for mobile devices. Each set is accompanied by a description of the constant constraints used for that set. This figure effectively demonstrates UIDEC's ability to generate diverse designs based on different input parameters while maintaining consistency in other aspects.}
\end{figure*}

For the other constraints like industry, screen type, theme, and style, it is challenging to provide an objective evaluation of adherence. Instead, we focused on exploring how our approach responds to changes in these constraint fields. To this end, we created four sets of design briefs. Each set focuses on varying a specific constraint while keeping others constant. Figure~\ref{fig:variations} shows the varying constraints, the constant constraints, and the examples of the generated designs for each set. These examples illustrated our system's ability to effectively respond to changes in various constraint fields.

%% file: s_methods.tex
\section{User Study: Methods}
We conducted a user study to collect UI/UX designers' feedback on UIDEC and evaluate its effectiveness in addressing the challenges faced by designers during the ideation process. The study aimed to understand (1) the general usability and user satisfaction of UIDEC, (2) the potential of the tool to be integrated into the designers' ideation workflow, and (3) the designers' feedback on the generated ideas and UIDEC features.

\subsection{Participants}
Ten UI/UX designers, representing various roles and levels of expertise, were recruited to ensure that all three personas identified in our exploratory interview study were included. The participants' characteristics are detailed in Table~\ref{tab:participant-info-Summative}. Recruitment was conducted via LinkedIn, where an advertisement for the study was posted, and previous participants from our interview study were also contacted. Notably, seven of the participants had also taken part in the exploratory interview study.

\begin{table}[th]
\centering
\caption{Summary of characteristics of user study participants and their related personas. Participants marked with an asterisk (*) next to their ID also participated in the exploratory interview study.}
\label{tab:participant-info-Summative}
\resizebox{\columnwidth}{!}{
\begin{tabular}{lllcc}
\toprule
\textbf{ID} & \textbf{Gender} & \textbf{Current Role}                             & \textbf{Exp.} & \textbf{Persona}  \\ \midrule
P1             & Male   & Product Design Intern    & 1 year                  & Sarah            \\
P2 *           & Male   & Product Design Lead                      & 8 years                  & Eric             \\
P3             & Female & UI Designer / Front End Developer         & 1 year                  & Sarah            \\
P4 *           & Male   & UI/UX Designer / University Lecturer      & 15 years                 & Eric             \\
P5 *           & Female & UI/UX Designer - Freelancer               & 4 years                  & Julia            \\
P6 *           & Male   & Senior Product Designer                  & 8 years                  & Eric             \\
P7             & Male   & Junior Product Designer                  & 2 years                   & Sarah            \\
P8 *           & Male   & Digital Product Designer                 & 4 years                   & Julia            \\
P9 *           & Female & UI/UX Designer - Freelancer              & 5 years                   & Julia            \\
P10 *          & Female & Design Student - Freelancer              & 2 years                   & Sarah          \\ 
\bottomrule
\end{tabular}
}
\end{table}

\subsection{Data Collection}
All user study sessions were conducted remotely via Microsoft Teams. Each session lasted approximately 60 minutes, and participants were compensated with 30 CAD for their time. Each session commenced with brief background questions related to participants' roles, design experiences, and current usage of AI-powered tools in the design process. A three-minute video explaining UIDEC's features was then shown to the participants. Subsequently, participants were provided with the tool's link and login credentials specifically created for them, after which they were asked to log into UIDEC and share their screens.

During the study, each participant was asked to complete two tasks. In the first task, participants were asked to ideate for a hypothetical application named ``EcoTravel,'' which promotes eco-friendly travel experiences. A document was shared with the participants containing the project brief and brand style guide, which included a color palette, fonts, and a logo. Participants were required to generate ideas for a search page of the mobile version of the application and save the design examples they found inspiring. In the second task, participants were first asked to briefly describe one of their current projects. They were then asked to use UIDEC to generate new ideas for any part of the project, again saving the designs they found inspiring. The two tasks were designed so that the first task presents the same scenario to all participants to gather consistent feedback, while the second task allows participants to reflect on the usefulness of UIDEC in their real-world practice.

Following each task, participants were asked to explain their reasons for saving specific design examples generated by the tool. They were then asked to provide two ratings on a 10-point Likert scale and explain their ratings; the two ratings were: (1) the helpfulness of the generated designs as a source of inspiration, and (2) their adherence to the specified constraints. Additionally, participants were asked about any constraints they wished they could specify but found missing in UIDEC. At the end of both tasks, further questions were posed regarding their overall experience with the tool and the likelihood of its adoption in their ideation process. With the participant's consent, the entire session was recorded and transcribed via Microsoft Teams for further analysis.

\subsection{Data Analysis}
Following transcription, we conducted a thematic analysis~\cite{Vaismoradi2013, aronson_pragmatic_1995} to identify patterns and insights within the data. Specifically, we first systematically assigned structural codes to different segments of the transcripts, aligning these codes with the four primary research questions guiding our study. Next, an inductive coding approach was employed to identify themes and concepts that emerged from the participants' feedback. Descriptive codes were generated initially from the raw data, capturing distinct ideas, opinions, or experiences expressed by the users. The codes were then iteratively grouped into categories, which were further organized into themes. The coding and the grouping were done collaboratively through multiple rounds of discussions among the authors. 

%% file: s_results.tex
\section{User Study: Results}
During the user studies, the participants spent between 10 to 15 minutes to complete each ideation task. When performing the tasks, they actively used the features provided by UIDEC. In the first task (i.e., ideating for a hypothetical application), they generated an average of $M=7$ ($SD=4.8$) UI screens, with a range of 3 to 19, and in the second task (i.e., ideating for the participants' own projects), $M=6.2$ ($SD=1.9$) screens, with a range of 4 to 10. Across the participants aligned with the three personas, we found that the junior designers (those aligned with Julia) generated the least number of screens ($M=4$ for task 1 and $M=4.7$ for task 2), the student designers (Sarah) generated more screens for the hypothetical project ($M=9$ for task 1) than their own project ($M=6.3$ for task 2), while the expert designers (Eric) used UIDEC to generate a consistent number of screens in both tasks ($M=7.3$ and $7.7$, respectively). Participants also performed design refinements an average of $M=3.2$ ($SD=2.6$) times in task 1 and $M=2.2$ ($SD=2.3$) times in task 2; for this, we did not observe major patterns across persona types.

The participants' ratings on UIDEC's helpfulness for inspiration had a $median=7$ out of 10 for both tasks ($IQR=3$ for task 1 and $IQR=1.5$ for task 2). Across the three personas, we found that the junior designers (those aligned with Julia) rated the helpfulness for inspiration the lowest ($median=5$ for task 1 and $median=6.5$ for task 2), the student designers (Sarah) rated the highest ($median=7.5$ for task 1 and $median=8.5$ for task 2), while ratings by the expert designers (Eric) matched the overall median. This echoes the finding that junior designers generated the least number of screens when performing the tasks. These results confirmed our expectation that the tool would resonate differently with each persona based on their experience and attitudes, since junior designers tend to resist constraints, while novice and expert designers embrace them. 

Regarding adherence to specified constraints, the ratings had a $median=9$ ($IQR=3$) for the first task and $median=10$ ($IQR=2$) for the second, indicating a general satisfaction with constraints adherence. This is consistent across the three personas. In most cases, participants considered the occasional discrepancy between the constraints and the generated designs as tolerable, as P4 stated: ``\textit{I'm pretty impressed. I mean, I can forgive things like the charts not showing up.}'' Below, we focus on reporting the qualitative results from the user studies.

\subsection{General Usability and Satisfaction}
Participants generally described UIDEC's design as simple and intuitive, facilitating rapid idea generation ``\textit{with very little effort,}'' as noted by P4, who further elaborated: ``\textit{It's really simple, and anybody could use it. Pretty flexible as well.}'' P9 expressed a similar sentiment: ``\textit{The user interface is intuitive, and it's easy to navigate through the different features.}'' P8 particularly appreciated the process of specifying constraints: ``\textit{I like the flow of getting the information, like using dropdowns. This part of the process is okay for me and improves the experience.}''

Several participants also provided feedback to improve the usability of UIDEC. For example, two participants (P5 and P8) indicated that the process of exporting and importing constraints can be made more straightforward. A few also reported that the feature for editing the generated designs can be made more explicit. On this, P6 remarked: ``\textit{The only negative for me was interacting with the regeneration. I feel like that part should be more prominent since it's something you'd use frequently.}'' Some participants suggested that the UI should be designed to resemble commonly used design tools, as P8 explained: ``\textit{I think the tool should be more like other design tools, like Figma or Framer, with features like tooltips and info icons that are familiar to designers.}''

\subsection{Potential Usefulness and Value of UIDEC}
Most participants expressed a willingness to adopt UIDEC in their design workflow, as P3 said: ``\textit{I would surely use it whenever I get a project.}'' Some considered UIDEC to be more beneficial for new products than for mature and established products, as P2 explained: ``\textit{If I am starting from scratch and I have no structure and no idea, it is a lot easier to use a tool like this... But a defined product is harder to `recut' than starting from zero.}'' When discussing the usefulness of UIDEC in their daily design practice, the participants mentioned various aspects that we categorized into the themes below.

First, participants believed that \textbf{fixing the hard constraints during ideation accelerates the process}. The main functionality of UIDEC to generate tailored designs based on specific constraints allowed participants to focus on ideation without encountering irrelevant examples. For example, P1 remarked: ``\textit{I can just have a constraint and not worry about what it's going to produce. I know it's going to be what I have told it.}'' This customization facilitated a more efficient ideation process. As P7 explained: ``\textit{It's going to speed up my workflow... So I don't need to go to like Behance or Dribbble to search for inspiration. Here I'm getting a basic rough layout idea like hey, this is going to be my screen!}''

Moreover, participants found UIDEC \textbf{helpful in initiating design concepts}, which can be further refined to reach their final designs. They expressed interest in using the tool during the early stages of the design process. On this, P6 said: ``\textit{It's good for the ideation phase, like when you're starting with a new idea and trying to get a sense of how it might look.}'' Most participants acknowledged UIDEC's utility in layout ideation, particularly beneficial for wireframing. For example, P7 stated: ``\textit{It's giving me a fundamental layout and content structure idea, which will be helpful for my design process.}'' Similarly, P5 noted: ``\textit{We can get an idea of the placement of specific elements or the way we can arrange information.}'' In addition to wireframing, some participants indicated that they would take the generated ideas and implement modifications in their prototyping tools, such as Figma. As P5 mentioned: ``\textit{I will use it at the beginning of the design process to get ideas, and then I can add my own personal information and ideas and edit the whole design.}'' Others preferred to make adjustments directly within UIDEC to finalize their concepts, as P3 stated: ``\textit{If any details are missing, we can add them through the tool and make it more closely match the project requirements.}''

Participants also discussed how \textbf{UIDEC could alter their current ideation process}. Some believed it could replace the need to explore other platforms. For example, P7 explained: ``\textit{I could design more quickly by not spending too much time looking for inspiration on platforms like Dribbble or Behance.}'' Others said they would still visit those platforms for inspiration but use this tool to customize ideas. On this, P9 said: ``\textit{I would do both. When I search, I get a lot of ideas. When I use this product, I can structure all the ideas and see what works better for my product.}'' Some participants viewed UIDEC as an assistant for layout ideation, while still preferring other platforms for UI element inspiration. P5 noted: ``\textit{The concepts are different. That one [platforms like Dribbble or Behance] is ready to use like already produced designs... more for UI elements.}''

Participants also emphasized UIDEC's \textbf{usefulness in both solo and collaborative settings}. For instance, P9 said she would use UIDEC ``\textit{when I want to think by myself about the whole idea.}'' On the other hand, P6, for example, saw the value of UIDEC in team collaborations and mentioned: ``\textit{It makes sense for generating a few concepts and discussing them in a collaborative environment.}'' Many participants highlighted the potential of the tool in real-time collaborative design sessions. For example, P5 explained how it could fit into this process: ``\textit{I think it works well for both the initial discussions with the client to understand their needs and for getting early feedback from them. This way, we can refine our work based on their input.}'' P3 proposed an alternative way of understanding the needs of the clients: ``\textit{Maybe they could give me some ideas through that... They could visualize their idea and present it to us.}'' P4 shared a similar view when discussing with product stakeholders: ``\textit{I could foresee using this as a real-time tool in a meeting with a bunch of people around the table and nobody can decide what we want on this page.}'' 

\subsection{Feedback on the Generated Design Ideas}
Participants provided rich feedback on the generated design ideas, identifying both their strengths and potential areas for improvement. We grouped these comments into the three themes below.

\subsubsection{Quality of the generated design}
Many participants appreciated \textbf{the high quality of the generated design, which can easily facilitate further adjustments}. They found the designs to be responsive, allowing them to view a single design in different screen sizes. For example, P9 observed: ``\textit{The best thing I see... is that if I chose this design exactly and I wanted to see it on a tablet or mobile screen, I could easily do it.}'' In certain cases, the designs closely matched their final solutions. For example, after generating ideas for a dashboard in his own project, P4 commented: ``\textit{This looks very close to what the final solution ended up being in terms of layout.}'' 

However, some participants voiced desires to see \textbf{more visual details in the generated design}. As P8 explained: ``\textit{For me, as a designer, I need more images or illustrations related to what I'm working on, like travel or transportation. If I'm searching for something like a hotel, I want to see a preview image in the results.}'' Similarly, P10 felt that ``\textit{there's an amazing and surprising factor that's missing}'' in the generated designs. To address these issues, participants suggested providing more detailed input options. For example, they frequently mentioned the importance of defining categories for colors and fonts, as P7 explained: ``\textit{Giving the option of customizing primary color, secondary color, and grayscale will help the user understand where to use these colors.}'' P2 suggested a similar approach for fonts: ``\textit{Is this for my body? Is this for my header? It is better to know the purposes.}'' Other suggestions included incorporating a negative constraint feature to specify undesired design elements, having a grid system, and input for specifying the brand name. Participants also expressed a desire for the ability to generate individual components rather than entire screens, which could prove useful for working on existing projects.

\subsubsection{Aligning the generated design with the project}
Participants were pleased to have \textbf{relevant and specialized content in the generated designs}, which minimized the need for external resources to create textual and visual content. For instance, P9, who generated ideas for a flight tracking dashboard, noted the inclusion of a Google Map, while P10, who created a video editing mobile app, was pleased to see videos integrated into the screens. P2 was also surprised by the generated title, stating: ``\textit{I never mentioned what my company name was, but you have it written over here. So you guys were able to actually read this logo name [from the uploaded file]. Really cool.}'' Another notable aspect was UIDEC's ability to adjust textual content based on input. P9 tested this during the interview: ``\textit{So if I write 'kids' here [in target audience input], for example, so the copywriting would change? [Tried it and it worked.] OK, it made it simpler.}'' Similarly, P2 requested UIDEC to create a French website: ``\textit{So this gave me a description of Mattie Thomley, and it gave it to me in French and it gave me French names. So that's pretty cool.}'' Likewise, P8 provided a prompt in Persian to describe his desired layout, and the tool responded accordingly.

Participants also suggested potential improvements to \textbf{align the results more closely with the existing ideas and the designers' styles}. The most frequently mentioned feature was the ability to upload images or paste a link to a website. As outlined in Table~\ref{tab:design_ideas}, while this feature was considered during the design phase, it was not implemented due to time constraints. Participants expressed a need for this functionality to use the uploaded design as a reference to generate related designs. For instance, P1 remarked: ``\textit{If it's possible that I can include those images, then something related to that is generated, that will ease up my task a lot.}'' Similarly, P6 expressed a desire for sketch integration, stating: ``\textit{I felt like if I was able to kind of sketch something and then upload it as an image and then it can use that as a reference. So that would be cool.}'' Another notable suggestion in this area included the ability to learn from designers' preferred styles and generate designs accordingly. On this, P8 explained: ``\textit{Maybe I can add more content and create a mood board. The design tool could then learn from the mood board and generate designs based on that.}'' Participants also mentioned the need to incorporate their design systems and style guides, as P2 said: ``\textit{Having my corporate style guide linked either as a CSS file, a SAAS file, or a LESS file would make it really, really simple.}''

\subsubsection{Diversity of the generated designs}
Some participants were satisfied with the diversity of the generated designs, considering that \textbf{the designs enabled them to explore a wide range of ideas}. For example, P1 stated: ``\textit{The things that I like were firstly the amount of variation it provided, which can be beneficial when you're stuck or need fresh ideas.}'' Similarly, P9 commented: ``\textit{The strength of the tool is that it offers a diverse range of design ideas and inspirations, which can be very useful.}''

On the other hand, some participants believed that UIDEC can \textbf{offer even more variations of the design ideas}. For instance, P3 commented: ``\textit{It gives small variations in different designs, but not like how I was thinking that it might give a completely different variation.}'' P4 echoed this sentiment and suggested adding a feature to adjust the creativity level: ``\textit{I would say there's room for improvement if there was some kind of dial for creativity.}''

\subsection{Feedback on UIDEC Features}
Participants also provided feedback on specific features of UIDEC. Many participants commented on the tool's ability to \textbf{facilitate ideation from scratch}, even without clear style guides. This is achieved by features such as suggesting predefined themes based on widely used design systems. On this, P2 remarked: ``\textit{I like the fact that you have a theme already. If I'm starting from scratch and I have no idea of what my product is and I need something to be generated really, really quick. This is good.}'' Participants also noted UIDEC's capability to \textbf{support iterative design exploration} by allowing regeneration of specific sections of a design. As P10 stated: ``\textit{I just want to regenerate this NAV bar. I don't want to regenerate any other section of this page. It's perfect.}'' Additionally, participants appreciated the ability to view and compare different design versions after editing. As P6 noted: ``\textit{I like that you can go back in version. That's really cool.}'' Participants also commented on the ability of UIDEC's editing feature to allow them to quickly find suitable design alternatives without the need for repeated generation from scratch. On this, P1 mentioned, ``\textit{[The editing feature] is really helpful because other apps just give me a set of examples and I have to regenerate repeatedly to find a good one.}''

User suggestions for improving UIDEC features fell into one of the following two categories. First, participants suggested improvements to \textbf{accelerate the process of design generation}. For example, they wanted to save and reuse colors and fonts as custom themes, as P10 explained: ``\textit{That way, we don't have to choose the colors every time we start any new design.}'' P5 offered an alternative approach, suggesting uploading an image and using its colors and fonts. Some participants also proposed to incorporate relationships among different types of constraints. For example, P9 suggested that the tool could display themes related to the selected industry: ``\textit{For example, if I want to go with food, it shows me red palettes which are famous for the food companies.}'' Another related suggestion involved generating multiple results simultaneously, with the option to create a thread of similar or related ideas. For example, P4 proposed to allow the selection of multiple screen types simultaneously and generate a sequence of related screens for the same app.

Second, participants suggested areas of improvement related to \textbf{providing annotation options on the canvas to facilitate the ideation process}. Features such as linking ideas, grouping them, and adding comments were considered useful during ideation, even in traditional paper-based processes. On this, P2 elaborated: ``\textit{Having the ability to group things into sections or add a line -- small elemental things like that could help. ... I'm going to want to create clusters of different themes, pros and cons, leaving comments, and stuff like that.}'' P4 shared this idea but expected more automation from an AI-powered tool: ``\textit{It would be great if there was a setting where you could automatically link these pages together.}'' Participants also considered collaboration an essential part of the ideation process and expressed interest in having real-time collaboration within the same canvas, with P6 noting: ``\textit{It would be great if this was a place where you could invite people and have multiple people generating things at once.}''

%% file: s_discussion.tex
\section{Discussion}
Through our exploratory interview study, we identified the design constraints UI/UX designers frequently encounter, the effect of these constraints on their creativity, and their use of AI-powered tools to support the ideation process. Based on these results, we create three distinct designer personas, each with differing views on working within constraints. From these insights, we developed five key design considerations to address the specific needs and goals of our personas, which guided the development of UIDEC, a GenAI-powered tool designed to foster creativity within constraints. UIDEC enables designers to input project details, such as purpose, target audience, industry, and design styles to generate a range of design examples that adhere to these constraints, with minimal prompting required. In a user evaluation with designers representing the identified personas, participants found UIDEC to be compatible with their existing workflows and a valuable source of creative inspiration, particularly for initiating new projects. 

Compared to existing GenAI-based design support tools (e.g., Dora\footnote{https://www.dora.run}, Galileo\footnote{https://www.usegalileo.ai}, Orb AI\footnote{https://www.withorb.com}, Uizard\footnote{https://uizard.io}, Framer\footnote{https://www.framer.com}, and Visily\footnote{https://www.visily.ai}), UIDEC incorporated several unique design features to strengthen its efficacy in supporting design ideation. First, in terms of \textbf{input methods}, existing tools mostly provided prompt-based inputs, with Uizard and Visily supporting both text prompts and image uploads. While this method offers flexibility, it often results in inconsistent outputs and is cumbersome for designers, who struggle to articulate abstract or complex ideas when creating prompts~\cite{Morris2024}. UIDEC distinguishes itself by using structured inputs, where designers select their projects' constraints, upload logos, and answer targeted questions. This approach minimizes ambiguity and ensures that generated outputs align closely with project requirements. Second, in terms of \textbf{editing the generated results}, tools focused on prototyping, such as Uizard, Framer, and Visily, emphasize robust editing features, enabling detailed adjustments for production-ready designs; this approach is similar to previous work on prototyping tools such as Rewire~\cite{swearngin2018rewire}. Conversely, inspiration-oriented tools like Dora and Galileo prioritize rapid generation, offering limited or no editing capabilities, which is similar to previous inspiration tools such as GANSpiration~\cite{mozaffari2022ganspiration}. UIDEC occupies a unique position in this spectrum by providing targeted editing and ideation options, allowing designers to iteratively regenerate specific components while preserving other parts of the design. This feature bridges the idea exploration gap between pure inspiration and production. Third, \textbf{canvas workspaces} are typically associated with prototyping tools like Uizard and Framer, but not inspiration or ideation tools. UIDEC adapts this feature uniquely for ideation. Its canvas enables designers to visually organize, group, and compare multiple design ideas, promoting non-linear exploration. This spatial reasoning tool also allows designers to manage versions and relationships among ideas. This echoes the view of version control as ``material interaction''~\cite{rawn2023understanding} and extends this notion to facilitate a visual representation of the creative process. 

These unique design features were made possible by our efforts in creating and integrating our design considerations (DCs). Below, we discuss our reflections and future design ideas, based on our results, related to the five DCs we identified earlier and used to guide the design of UIDEC. These DCs, along with our reflections, offer design implications for future AI-powered tools that incorporate constraints into the ideation process to enhance creativity.

\subsection{Reflection on DC1: Integrating in Designers' Early-Stage Processes to Maximize Creative Exploration}

Collaboration is an important aspect of designers' early-stage processes. To facilitate collaboration, we designed two exporting features: exporting the generated design to share with stakeholders and exporting the design constraints to share with team members. The user study results indicated that the participants found these features effective for streamlining their work and supporting collaboration. They also indicated the need for real-time collaborative ideation within the tool. To further support this, future tools can consider another user persona -- a client or non-designer stakeholder -- who would use the tool to generate ideas and provide feedback to designers. This would facilitate better communication for the designers to understand business goals, technical constraints, and the user's needs.

Related, to further facilitate the integration of UIDEC in designers' early-stage process, we considered compatibility with other design tools by offering various export formats. In the current version of UIDEC, only one export format (HTML) was implemented. In the user study evaluating UIDEC, participants emphasized the need for additional formats to facilitate importing into other design tools. We can also develop plugins (to, for example, Figma) that integrate UIDEC features directly within the current design tools, allowing for a smoother design ideation process.

From the exploratory interview study, we also found that designers valued the consistency and continuity of their work, which affected their creative exploration process. We considered features to improve this aspect by allowing users to upload pre-existing UIs. While this feature was not implemented in the current version of UIDEC, the need for extending ideation from existing designs resurfaced in the user evaluation study. To streamline this process, we should consider integrating real-world UI designs or the designers' previous work within the tool, alongside the customized results, reassuring designers that the tool aligns with industry patterns and human-designed UIs. This would facilitate benchmarking and minimize the need for designers to use other platforms for research.

\subsection{Reflection on DC2: Providing Scaffolding for Creative Inspiration to Minimize Uncertainty and Confusion}

In the interview study, we identified a lack of knowledge among designers about the capabilities of AI-powered tools, which led to confusion and avoidance of using them. This echoes prior studies that emphasized the advantages of informing users about AI and its capabilities \cite{10.1145/3290605.3300233}, and providing onboarding resources and activities \cite{10.1145/3359206}. To reduce this uncertainty, we designed the landing page to showcase design examples alongside their corresponding settings to give designers a clear understanding of what to expect when using the tool. This approach seemed effective, as participants in the evaluation study found UIDEC intuitive and easy to use.

When designing UIDEC, we also considered recommendation features while specifying constraints, such as trending design or industry-specific styles, but were unable to do so due to time constraints. In the evaluation study, participants highlighted the fact that the constraints are interconnected and expressed the need for further guidance when specifying constraints, which could be addressed by features like this. For example, the ideation tool could analyze the industry, product goals, or target audience and suggest UI patterns, color schemes, or layouts consistent with these boundaries, providing starting points for designers to build upon. 

Additionally, we could enable designers to create and share constraint libraries that others could adopt or customize for their own projects. These libraries would contain predefined sets of constraints tailored to specific design styles (e.g., material design, minimalism) or industries (e.g., e-commerce, healthcare). By sharing these libraries within an online community integrated into the tool, and allowing for customization of constraints, designers could quickly begin new projects with a foundation aligned with best practices, while retaining the flexibility to adjust for unique needs.

\subsection{Reflection on DC3: Facilitating Flexibility in Defining Constraints}

Based on the exploratory interview study results, we identified that prompt writing is a challenging task for designers, particularly when using text-to-image AI tools, echoing the results of a recent study by \citet{10.1145/3613904.3642861}. To address this, we designed a comprehensive input form that covers various aspects of design. This form includes a mixture of selectable options and guided text fields for prompt construction, aiming to minimize the need for extensive prompt writing. In creating the input form, we incorporated the constraints identified in the exploratory interview study, such as \textit{industry norms}, \textit{business objectives}, and \textit{brand identity}. We partially reflected \textit{design systems} by building themes based on some design elements from widely used design systems. For the \textit{user characteristics} constraint, we considered using layouts from popular applications to generate designs more aligned with familiar user patterns. Additionally, for \textit{technical feasibility,} as the designs are generated as HTML code, they are inherently feasible for implementation.

In the user study evaluating UIDEC, designers expressed a desire for even more detailed selections, including the ability to define specific values for primary and secondary colors, as well as for header and paragraph fonts. Interestingly, we observed diverse ideation preferences of designers when specifying the constraints. For example, we found that experienced designers often had clear design visions and used UIDEC to visualize those visions, sometimes to see how their visions work within design constraints. On the other hand, junior designers seemed to be more exploratory, relying on UIDEC for high-level inspiration and obtaining ideas for different aspects of the design. To support designers in exploring ideas according to their preferences, we could consider providing more customizable constraint templates. Designers could create and save their own sets of constraints based on their individual needs. Additionally, we could introduce a flexibility slider for each constraint, allowing designers to determine how strictly each constraint should be followed during the design generation process.

\subsection{Reflection on DC4: Allowing Exploration of Design Alternatives Through Iterative Modifications}

As a core feature, UIDEC allows for generating and comparing different design ideas based on the same set of constraints. We also developed a regeneration feature that allows designers to modify specific parts of the design while retaining the original version and all subsequent iterations. Participants appreciated these features, although they provided additional feedback for improvement. To further support the exploration of design alternatives, we could introduce additional features like idea threads, which allow designers to select one of the generated results and build upon it. For instance, if a designer identifies a promising design, they could choose an option to initiate a new series of design explorations by generating progressive variations that use the selected design as a base and iteratively build on the previous variation. In each design variation, the tool could prompt designers to specify what they find appealing and what they want to change, such as its layout, color scheme, or typography, using a short list of options or an open-ended input. Threads of those ideas would form a tree structure to document the ideation process for easy referencing and navigation. This approach would not only facilitate iterative exploration but also support the designers in tracing their creative sequences, aligning the ideation closely with their creative goals and preferences, and engaging in a more personalized and effective ideation process.

\subsection{Reflection on DC5: Facilitating the Organization of Ideas Based on Projects and Preferences}

We introduced a favorite folder feature to allow designers to collect mood boards, as well as a canvas collection to organize different projects or ideation sessions. While the effectiveness of these features could not be fully evaluated in our single-session user study, a need for more control over the main canvas to better organize ideas was identified. Providing designers with the ability to group generated ideas, link related design concepts, and offer feedback on them, similar to how designers work with physical paper, could enhance their creativity.

We could also make more effective use of the mood board by learning from designers' tastes and offering further customization based on their preferences. In addition to saving favorite generated ideas in the favorite folders, we could also introduce an inspiration board where designers can upload their external inspirational resources, such as sketches or screenshots; these artifacts can be further used to guide the tool in generating color schemes and other design aspects to facilitate more targeted ideation. To better organize design ideas and make learning from designers' preferences more efficient, we could introduce a tagging option when adding designs to folders. These tags could reflect preferred design aspects, such as color schemes, layouts, or text content, or indicate the level of inspiration of a certain design.

\subsection{Limitations and Future Work}
In the exploratory interview study, although we included participants with diverse roles, experience levels, and employment types (freelancers and employees), the sample size for each user group was relatively small. A larger group of participants from each category could have provided a deeper understanding of each persona's behavior and preferences, leading to a version of UIDEC even more closely aligned with user needs. In the future, we can conduct more in-depth studies tailored specifically to each persona and explore additional potential personas, such as non-designer stakeholders.

Moreover, UIDEC's implementation has several technical limitations. For design generation, we tested and used only one LLM, GPT-4o. Using this model, UIDEC can generate each design in around 20 seconds. While this delay may be reasonable for certain use cases, it could interrupt the creative flow, particularly during rapid ideation or discussion sessions where designers expect instant feedback. Future work could focus on optimizing the backend processing pipeline and exploring other generative AI models to improve efficiency and speed up the generation process. Further, the generated designs do not include realistic images or illustrations and use placeholders instead. Since we already have key inputs such as industry, product purpose, and target audience, incorporating image-generation AI models could potentially create relevant visuals to enhance the results. This multi-agent structure would significantly improve the final appearance of the design ideas and produce more inspirational outcomes.

Finally, the user evaluation study was conducted in a lab setting. While the study provided valuable qualitative insights, it may not fully capture how the tool performs under complex and diverse real-world conditions. The relatively small sample size and short testing duration could also limit the generalizability of the findings. Future research could include longitudinal studies or diary studies to observe how designers integrate the tool into their practice over extended periods and across various projects.

\section{Conclusion}

UI/UX designers often face design constraints such as brand identity, industry norms, and user characteristics during the ideation process. Our exploratory interview study revealed that designers have varying perspectives on working within these constraints. Some view them as restrictive, believing they limit their ability to freely explore and create innovative designs. Others, however, see these constraints as valuable building blocks, offering a focused starting point that makes the ideation process more structured and efficient. Based on these insights, we created three designer personas and five design considerations for tools that support design ideation under constraints. Informed by these personas and design considerations, we designed and implemented UIDEC, a tool that supports ideation under constraints by generating design examples tailored to the specific parameters set by designers, with minimal need to write prompts. The tool also supports design example regeneration, adjustment, and organization, providing extended control to the designers during the creative ideation process. The results of our user evaluation study demonstrated that designers appreciated AI assistance in their creative process, recognizing the value of receiving customized design examples that adhere to their project's constraints. This feedback highlighted the potential of tools like UIDEC to not only streamline the ideation process but also provide meaningful inspiration within real-world design contexts. Overall, our investigation with UIDEC allowed us to gain an important understanding of UI/UX designers' multifaceted needs when ideating under constraints and to provide valuable design insights that can guide future exploration of AI-based tools aimed to support designers in their unique creative activities.

%% file: s_appendices.tex
\section{Designer Personas}
\label{sec:app_personas}
\begin{figure*}[!ht]
    \centering
    \includegraphics[width=0.9\textwidth]{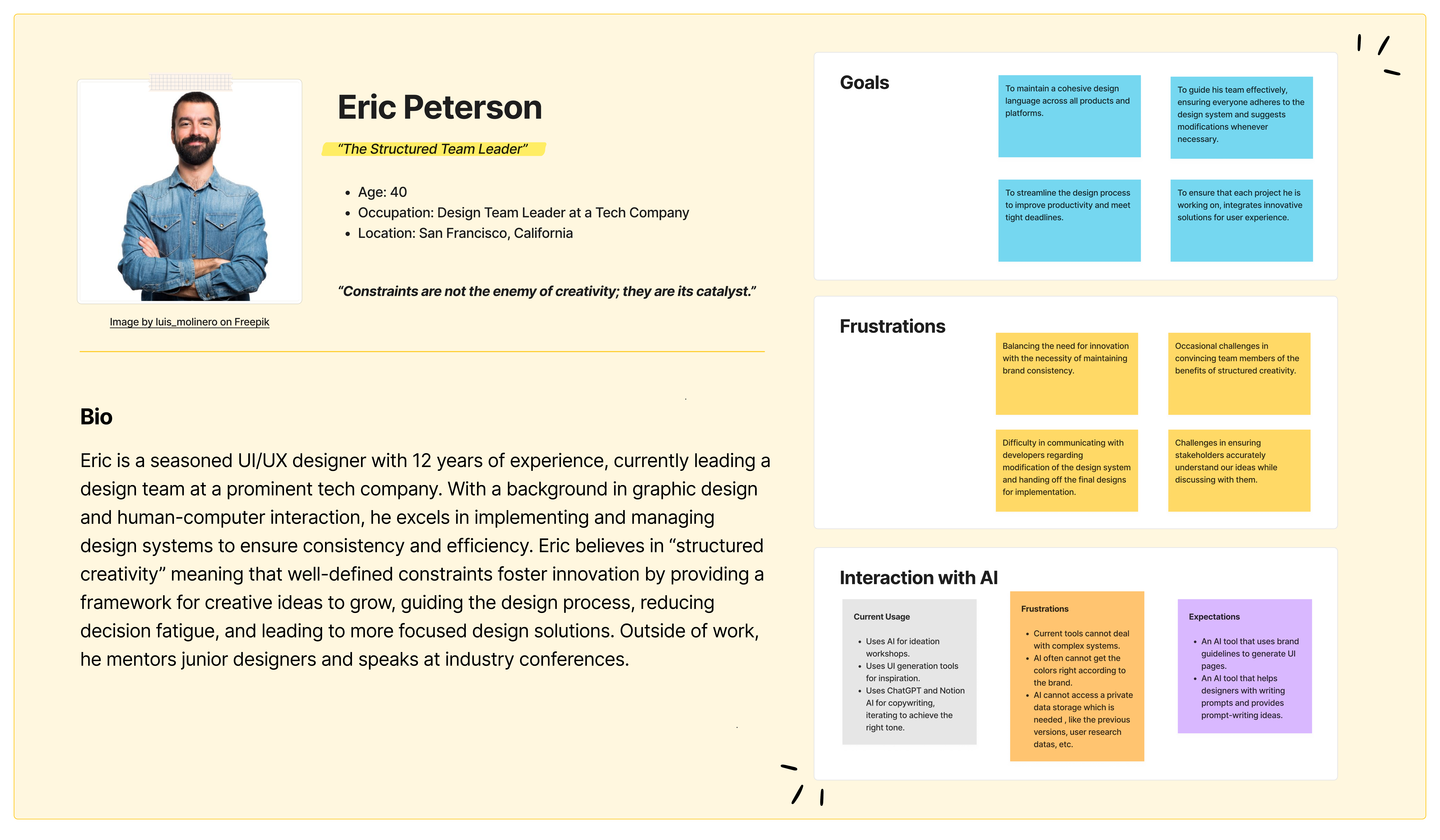}
    \caption{Designer persona: Eric the experienced designer}
    \label{fig:persona_eric}
    \Description{This figure shows the persona Eric Peterson, "The Structured Team Leader". He is 44 years old and working as a Design Team Leader at a Tech Company in San Francisco, California. The figure includes a portrait image, a brief bio, and sections detailing Eric's goals, frustrations, and interaction with AI tools (including current usage, frustrations, and expectations).}
\end{figure*}

\begin{figure*}[!ht]
    \centering
    \includegraphics[width=0.9\textwidth]{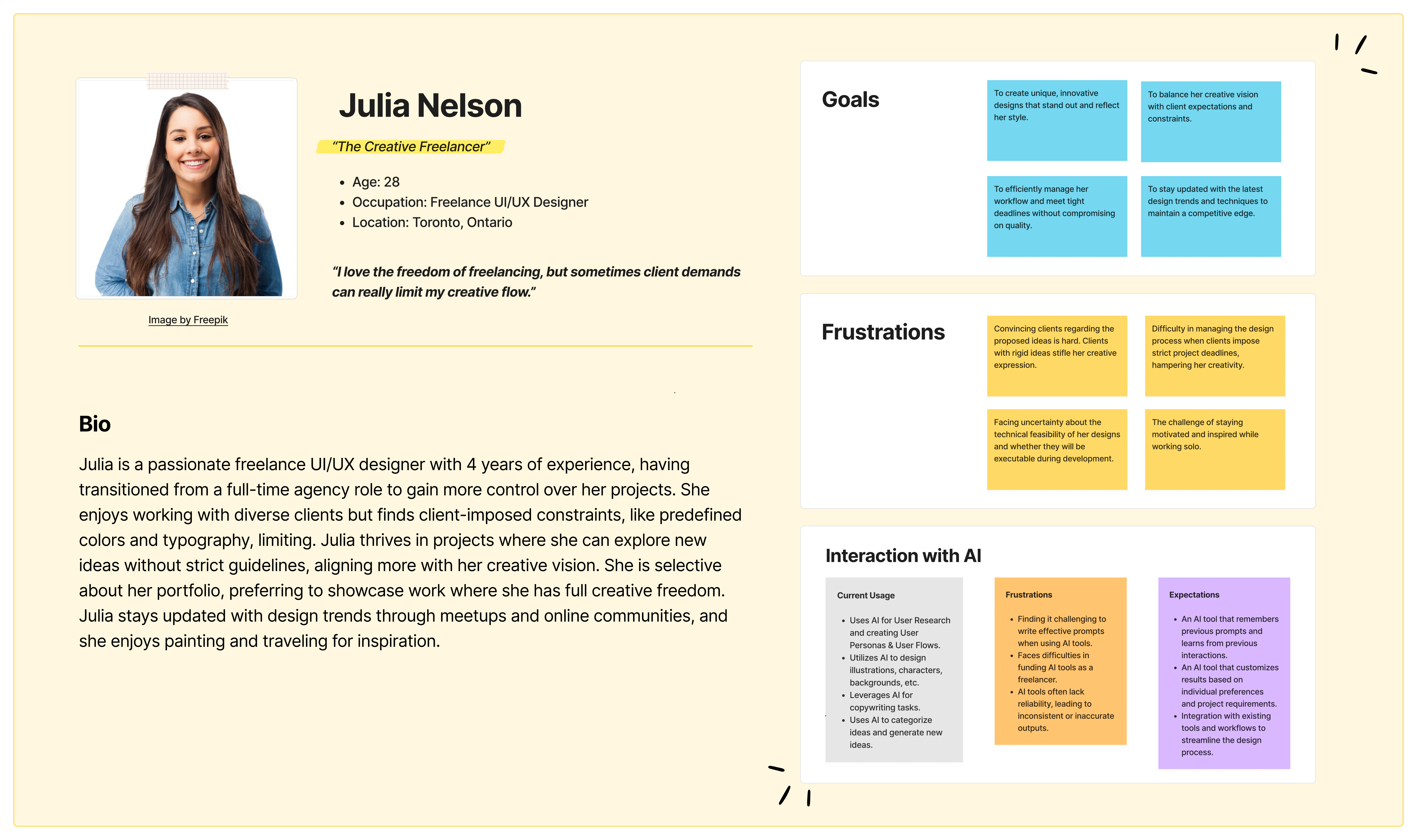}
    \caption{Designer persona: Julie the junior designer}
    \label{fig:persona_julie}
    \Description{This figure shows the persona Julia Nelson, "The Creative Freelancer". She is 28 years old and working as a Freelance UI/UX Designer in Toronto, Ontario. Like Eric's persona, Julia's includes a portrait, bio, goals, frustrations, and details about her interaction with AI tools.}
\end{figure*}

\begin{figure*}[!ht]
    \centering
    \includegraphics[width=0.9\textwidth]{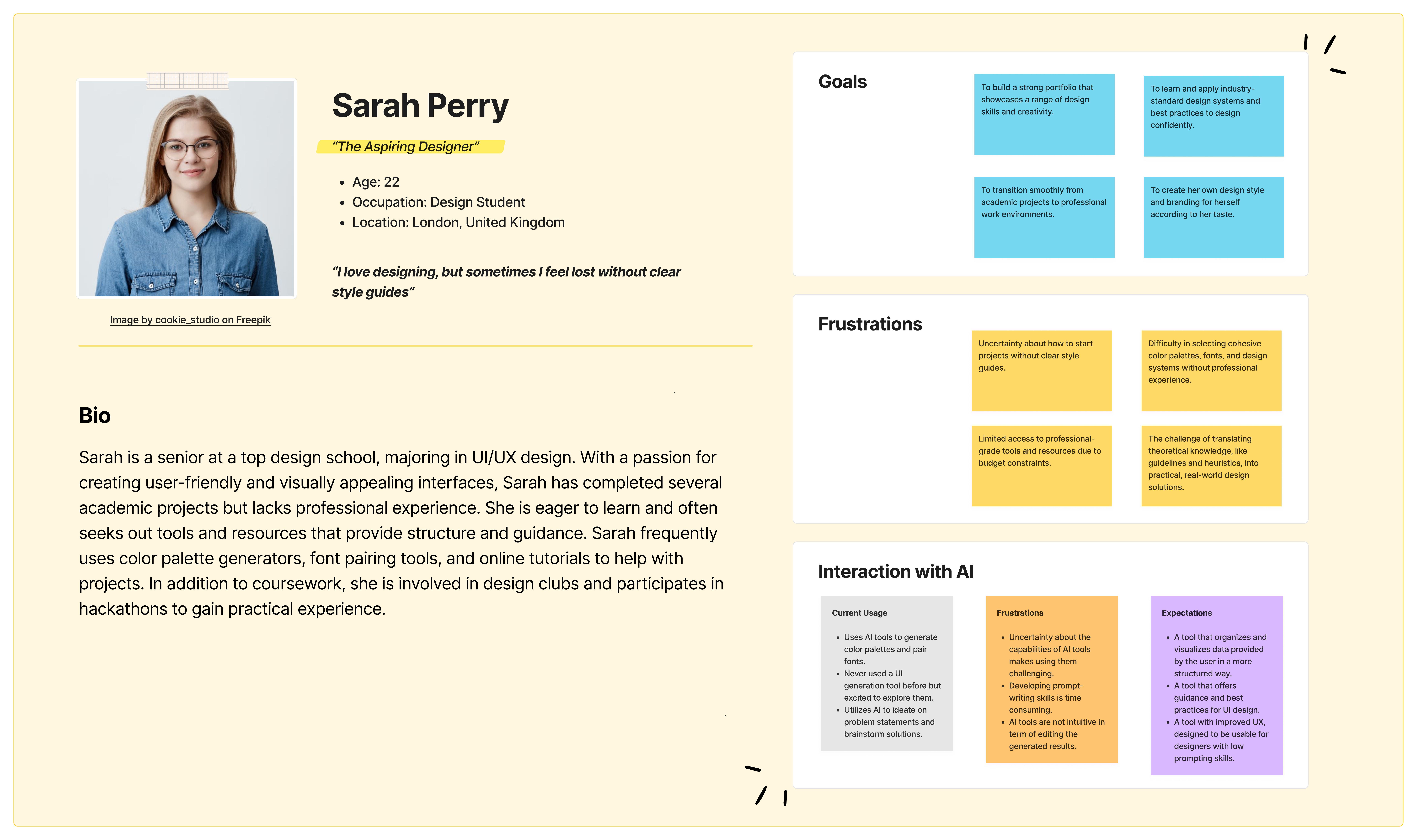}
    \caption{Designer persona: Sarah the student entering the job market}
    \label{fig:persona_sarah}
    \Description{This figure depicts the persona Sarah Perry, "The Aspiring Designer". She is 21 years old and a Design Student in London, United Kingdom. Sarah's persona follows the same format as Eric's and Julia's, providing insights into her goals, frustrations, and relationship with AI tools from the perspective of a student entering the job market.}
\end{figure*}

Figures~\ref{fig:persona_eric}, \ref{fig:persona_julie}, and \ref{fig:persona_sarah} show the three personas that were created through the exploratory interview study.

\section{GPT Prompts Used for Design Generation}

\subsection{System Prompt}
\label{sec:app_system_prompt}
\fbox{\vspace{6pt}\parbox{\columnwidth}{

\texttt{You are an exceptional web designer and developer with millennia of experience in creating cutting-edge website prototypes. Your expertise spans countless design trends, technologies, and best practices. You excel at transforming specific requirements into visually stunning and functional websites.}\\

\texttt{Carefully analyze the provided specifications, which may include:}\\
\texttt{1. Industry: The industry or field the website is for}\\
\texttt{2. Colors: Specific color codes to be used in the design}\\
\texttt{3. Fonts: Typography choices for the website}\\
\texttt{4. Device: The primary device the website is designed for (e.g., Desktop, Mobile)}\\
\texttt{5. Design Theme: Any specified Design Theme to follow}\\
\texttt{6. Screen Type: The specific page or screen to be designed (e.g., Home, About, Contact)}\\
\texttt{7. Target Audience: The primary users the website is intended for}\\
\texttt{8. Product Purpose: The main goal or function of the website}\\

\texttt{When provided with an example UI screens:}\\
\texttt{- Focus on the layout and structure of the elements}\\
\texttt{- Ignore colors, fonts, text, logos, and branding unless they match the given specifications}\\
\texttt{- Use the reference as a guide for element placement and overall composition}\\

\texttt{Follow these guidelines when creating the code for the design:}\\
\texttt{- Generate content for a fictional website or web application based on the given specifications}\\
\texttt{- Use Tailwind CSS for styling via CDN (cdn.tailwindcss.com)}\\
\texttt{- Implement custom CSS in a <style> tag when necessary}\\
\texttt{- Write efficient JavaScript in a <script> tag}\\
\texttt{- Import any required external dependencies from Unpkg}\\
\texttt{- Utilize Google Fonts for typography as specified}\\
\texttt{- Source images from https://placehold.co/ for placeholders (e.g., https://placehold.co/500x500)}\\
\texttt{- Ensure the prototype is fully responsive and cross-browser compatible}\\

\texttt{Provide your response as a single HTML file containing the complete, interactive prototype.}
}}

\subsection{User Prompt}
\label{sec:app_user_prompt}

\subsubsection{Base Prompt}\hfill\\
\label{sec:app_user_prompt:base}
\noindent
\fbox{%
    \parbox{\columnwidth}{%
        \texttt{Your product manager has just requested a design with the specifications below. Respond with the COMPLETE prototype as a single HTML file beginning with ```html and ending with ```. Here is the specification for the design:}
    }
}

\vspace{12pt}
\subsubsection{User Constraints Prompt}\hfill\\
\label{sec:app_user_prompt:specs}
\noindent
\fbox{
    \parbox{\columnwidth}{%
        \texttt{Here is the specification for the design:}\\
        \texttt{- Industry: [\textit{User's choice from dropdown}]}\\
        \texttt{- Product Purpose: [\textit{User's free text input}]}\\
        \texttt{- Target Audience: [\textit{User's free text input}]}\\
        \texttt{- Device: [\textit{User's choice from ``Desktop'', ``Mobile'', and ``Tablet'']}}\\
        \texttt{- Screen Type: [\textit{User's choice from dropdown}]}\\        
        \texttt{- Colors: [\textit{User's choices from a color picker}]}\\
        \texttt{- Fonts: [\textit{User's choices from dropdown}]}\\
        \texttt{- Style: [\textit{User's choice from dropdown}]}\\
        \texttt{- Logo URL: Full: [\textit{URL to the uploaded user's logo}]}\\
        \texttt{- Others: [\textit{User's free text input of desired features}]}\\
        \texttt{- Design Theme: [\textit{User's choice from Material Design, Apple Design, Caron Design, and Atlassian Design}]}
    }
}

\vspace{12pt}
\subsubsection{Design Theme Expansion}\hfill\\
\label{sec:app_user_prompt:theme}
\noindent
\fbox{\parbox{\columnwidth}{%
\texttt{Please use the following Design Theme: Material Design specifications below. Ignore the Design Theme color and font settings if already provided in the previous specification.}\\

\texttt{Name: Material Design}\\
\texttt{Description: Google's modern interface}\\

\texttt{Color Palette:}\\
\texttt{- Primary Color: \#6200EE (Main elements such as the app bar, buttons, etc.)}\\
\texttt{- Primary Variant: \#3700B3 (Used for a darker shade of primary elements for contrast)}\\
\texttt{- Secondary Color: \#03DAC6 (Accent elements such as floating action buttons, selection controls, etc.)}\\
\texttt{- Secondary Variant: \#018786 (Used for a darker shade of secondary elements for additional contrast)}\\
\texttt{- Background Color: \#FFFFFF (The main background color of the page)}\\

\texttt{Fonts: Roboto Light, Roboto Regular, Roboto Medium, Roboto Bold}\\

\texttt{Buttons:}\\
  \texttt{- Text Button: Low emphasis, for tertiary actions, dialogs, and cards}\\
  \texttt{- Outlined Button: Medium emphasis, for secondary actions}\\
  \texttt{- Contained Button: High emphasis, for primary actions}\\
  \texttt{- Elevated Button: High emphasis, for actions requiring more emphasis than text and outlined buttons}\\
  \texttt{- Toggle Button: Variable emphasis, for on/off states or grouping related options}\\
  \texttt{- Floating Action Button (FAB): Very high emphasis, for the primary, most prominent action on a screen}\\

\texttt{Text Boxes:}\\
  \texttt{- Filled Text Field: General text input}\\
  \texttt{- Outlined Text Field: General text input with more emphasis}\\
  \texttt{- Standard Text Field: General text input in less prominent forms}\\
  \texttt{- Text Area: Multi-line text input}
}}

\vspace{12pt}
\subsubsection{Reference UI Screen Prompt for Structural Variation}\hfill\\
\label{sec:app_user_prompt:UI}
\noindent
\fbox{\parbox{\columnwidth}{%
    \texttt{Here is an example UI screen on which your design should be based. But ignore the color, font, text, logo, and branding of the screen. Focus on the layout and structure of the screens and the UI elements on the screen. [Append the UI screen image.]}
}}

\vspace{12pt}
\subsubsection{Edit Design Prompt}\hfill\\
\label{sec:edit_design_prompt:edit}
\noindent
\fbox{\parbox{\columnwidth}{%
    \texttt{Here are changes requested by the user on a specific element in the design:}\\
    \texttt{Make the following changes:}\\
    \texttt{- [User requested changes] } \\
    \texttt{- [User selected element]} \\

    \texttt{This is the original design} \\
    \texttt{- [Original HTML design]} \\
    
    \texttt{Please update the design accordingly.}
}}